\documentclass[onecolumn,sort&compress,numbers]{els-mrw} 

\usepackage{amsmath,amssymb,amsfonts,amsthm,makeidx,graphicx,exscale,cancel,empheq,slashed,epsfig,multicol,color,blindtext,fancyhdr,mathtools,floatrow}
\usepackage{txfonts}
\usepackage{helvet}
\usepackage{mathrsfs}

\DeclareMathOperator*{\SumInt}{%
\mathchoice%
  {\ooalign{$\displaystyle\sum$\cr\hidewidth$\displaystyle\int$\hidewidth\cr}}
  {\ooalign{\raisebox{.14\height}{\scalebox{.7}{$\textstyle\sum$}}\cr\hidewidth$\textstyle\int$\hidewidth\cr}}
  {\ooalign{\raisebox{.2\height}{\scalebox{.6}{$\scriptstyle\sum$}}\cr$\scriptstyle\int$\cr}}
  {\ooalign{\raisebox{.2\height}{\scalebox{.6}{$\scriptstyle\sum$}}\cr$\scriptstyle\int$\cr}}
}

\usepackage{amsmath}
\usepackage[colorlinks=true, linkcolor=blue, citecolor=blue
]{hyperref}

\def\hhref#1{\href{http://arxiv.org/abs/#1}{arXiv:#1}}

\def\be{\begin{equation}}
\def\ee{\end{equation}}
\def\ba{\begin{array} }
\newcommand{\Tr}{\,{\rm Tr}}
\def\bac{\begin{array} {c}}
\def\bacc{\begin{array} {cc}}
\def\baccc{\begin{array} {ccc}}
\def\bacccc{\begin{array} {cccc}}
\def\ea{\end{array}}
\def\bea{\begin{eqnarray}}
\def\eea{\end{eqnarray}}
\def\x{\textcolor{red}}
\definecolor{red}{rgb}{1,0,0}

\def\psl{\hbox{\hbox{${p}$}}\kern-1.9mm{\hbox{${/}$}}}
\def\dsl{\hbox{\hbox{${\partial}$}}\kern-2.2mm{\hbox{${/}$}}}
\def\Dsl{\hbox{\hbox{${D}$}}\kern-2.6mm{\hbox{${/}$}}}

\setcitestyle{authoryear,round}
\begin{document}

%{\bf\LARGE Thermal Field Theory for General Equilibrium Density Matrices}
\chapter{Thermal Field Theory for General Equilibrium Density Matrices}\label{chap1}
\vspace{1cm}

\author[1]{Alberto Salvio}%
%\author[2]{Second Author}%
%
%\author[1,2]{Third Author}%

\address[1]{\orgname{University of Rome Tor Vergata}, \orgdiv{Physics Department and INFN}, \orgaddress{via della Ricerca Scientifica, I-00133 Rome, Italy}}
%\address[2]{\orgname{Name of Institute}, \orgdiv{Division or Department}, \orgaddress{Address of Institute}}

%\articletag{Chapter Article tagline: update of previous edition,, reprint..}

\maketitle

\begin{glossary}[Keywords]
Quantum Mechanics, Special Relativity, Field Theory, Statistical Mechanics
\end{glossary}

\begin{abstract}[Abstract]
Thermal field theory (TFT) is the combination of statistical mechanics and relativistic quantum field theory.
This (book) chapter provides a pedagogical introduction to (and a review of)  TFT  for general equilibrium density matrices, which allow arbitrary values not only of temperature and chemical potentials, but also of average angular momentum. A general technique to compute ensemble averages is illustrated.  Path-integral methods are developed to study the partition function and the thermal Green's functions of general operators, with an arbitrary number of points, in all field theories of physical relevance.
\end{abstract}

% XXXXXX\section*{Objectives}
%\begin{itemize}
%	\item Introduce TFT for general equilibrium density matrices.
%	\item Cover all types of fields of physical interest.
%	\item Describe general free theories in simple terms.
%	\item Include interactions  in a model-independent way.
%\end{itemize}

%\mytoc
%\tableofcontents

\section{Introduction}\label{Introduction}

Statistical mechanics and (relativistic quantum) field theory were first combined in the non-relativistic setting in the late 1950s to theoretically describe condensed and nuclear matter in standard laboratory experiments. One of the most famous pioneering work of those years was provided by~\citet{Matsubara1955}, which showed how a finite-temperature description of field theory can be obtained by formally substituting time with an imaginary quantity. This approach is known as the imaginary-time formalism. In the fifties the work of~\citet{Kubo1957},~\cite{Martin1959} (KMS) showed the existence of  some consistency conditions in finite-temperature field theory (the KMS conditions), which later turned out to be very important in TFT.

\cite{Fradkin1965} pioneered the combination of statistical mechanics with the relativistic version of field theory (i.e.~TFT).  
Several years later \citet{Matsumoto1983} extended the work of~\citet{Matsubara1955} showing that a finite-temperature description of field theory can also be obtained by thinking time as a variable on a contour $C$ of the complex-time plane. Choosing $C$ to be the imaginary axis recovers the Matsubara approach, while other choices, which include the real axis, allow to obtain observables directly at real time. The latter is known as the real-time formalism. In the imaginary-time formalism TFT is described as a Euclidean field theory with periodic (for bosonic fields) or antiperiodic (for fermionic fields) boundary conditions with respect to the imaginary time. Therefore, the imaginary-time formalism is suited to perform lattice non-perturbative calculations. On the other-hand, the real-time formalism is convenient when time-related observables are computed.

In several works completed decades after the work of \cite{Fradkin1965} it was also realized that TFT is an essential tool to describe processes in the cosmological plasma: in that context  the presence of a huge number of particles prevents the exact (non-statistical) description and relativistic effects are important. For example, at some point during the history of the universe, when the temperature was sufficiently high, quarks and gluons behaved as weakly interacting particles, almost free to move in their plasma (the quark-gluon plasma) and TFT is an essential tool to describe such physical situation. Laboratory experiments, conducted for example at CERN, 
aim to reproduce the quark-gluon plasma through  collisions of heavy ions (e.g.~gold or lead nuclei) and  TFT is the theory whose predictions are compared with the results of these experiments. 

Furthermore, TFT can also describe some aspects of the physics of compact astrophysical objects. In these cases it is often required to include  a finite value not only of the temperature, but also of the density of some particle species. This is achieved in TFT by switching on some chemical potentials. One famous example is provided by neutron stars, which are characterized by finite densities of neutrons, protons, electrons and other particles. An introduction to TFT applications to neutron stars is given, for example, by~\cite{Vuorinen2024} (see also the references therein). 

More generally, there are several introductions to TFT and its applications in the literature and some of them (\citet{Landsman1987},~\citet{Quiros1994},~\citet{Das1997},~\citet{Nair2005},~\citet{Bellac2011},~\citet{Laine2016},~\citet{Ghiglieri2020}) have significantly helped the author understand some subjects discussed here and in a previous introduction to TFT~\citep{Salvio2025}, henceforth I1. The latter is a recent  introduction to TFT from first principles at the graduate-student level. In all previous introductions to equilibrium TFT the author is aware of (including I1)  the average angular momentum is not included in the field-theoretic description, although its presence is compatible with thermal equilibrium. Moreover, often the chemical potentials are switched off.

The present chapter is an extension  of I1 to include the average angular momentum and a general set of chemical potentials (to include general equilibrium density matrices).  We will refer to I1 several times to indicate how to use it in order to complement the discussions of this chapter.  

TFT based on general equilibrium density matrices  is motivated, for example, by neutron stars, which typically feature, as previously mentioned,  finite densities of various types of particles  and are often quickly rotating because of their small size (as theoretically anticipated by~\cite{Pacini1967} and confirmed by the discovery of pulsars). Moreover, other compact objects, such as ordinary and primordial black holes and exotic compact objects, which can also be quickly rotating, furnish further motivations. For example, the accretion disks and coronas around black holes can be often considered rotating plasmas in approximate thermodynamic equilibrium.  Moreover, one can conceive reproducing a TFT with average angular momentum %and chemical potentials 
in a lab, engineering a quantum relativistic rotating plasma. 

Like in I1 all types of physically relevant interactions and a general field content is covered here, including scalars, fermions (both of Dirac and Majorana type) and vectors. In other words a general (interacting) gauge theory coupled to an arbitrary matter sector is considered. In such general setup this chapter pedagogically introduces techniques to compute ensemble averages and path-integral methods to represent the partition function and describe interactions.

\begin{BoxTypeA}[chap1:box1]{This chapter is organized as follows.}
\begin{itemize}
\item Sec.~\ref{density matrix} introduces the general equilibrium density matrix.
\begin{itemize} 
\item Sec.~\ref{Rest frame} finds the rest frame of the plasma and how to convert the results in a generic inertial frame. 
\item Sec.~\ref{Ensemble averages} provides general expressions for the ensemble averages of key observables.
\end{itemize}
\item Sec.~\ref {Free fields} discusses free fields.
\begin{itemize} 
\item Sec.~\ref{Free Scalars} covers free scalars.
\item Sec.~\ref{Free Vectors} adapts the discussion of free scalars to free vectors.
\item Sec.~\ref{fermion Free fields} covers free fermions, both of Dirac and Majorana type.
\end{itemize}
\item Sec.~\ref{Interacting theories and general path-integral formulas} extends the previous discussion to interacting theories and derive path-integral formulas.
\begin{itemize}
\item Sec.~\ref{Generalized Kubo-Martin-Schwinger condition} discusses generalized KMS conditions.

- ~~Sec.~\ref{Coordinate space} derives the generalized KMS conditions in coordinate space.

- ~~Sec.~\ref{Analyticity} discusses an analyticity property of a 2-point function that appears in these conditions.

- ~~Sec.~\ref{Momentum space}  derives the generalized KMS conditions in momentum space.

\item Sec.~\ref{ScalarPI} derives the scalar path integral.

\item Sec.~\ref{ScalarProp} computes the scalar thermal propagators.

\item Sec.~\ref{FermionPI} derives the fermion path integral, both of Dirac and Majorana type.
\item Sec.~\ref{Fermion thermal propagator} computes the fermion thermal propagators, both of Dirac and Majorana type.
\item Sec.~\ref{GaugePI} derives the path integral in gauge theories.
\item Sec.~\ref{propGauge} computes the gauge thermal propagators, including both the gauge fields and the Faddeev-Popov ghosts.
\item Sec.~\ref{Interactions} discusses the effects of the average angular momentum on interactions in gauge TFTs, and its role in perturbation theory.
\item Sec.~\ref{Summary and conclusions} offers a summary and concluding remarks.
\end{itemize}
\end{itemize}
\end{BoxTypeA}

This chapter is based on three recent works by the author, where a systematic and general treatment of equilibrium TFT was provided for scalars~\citep{Salvio2025b}, fermions~\citep{Salvio2025c} and gauge fields~\citep{Salvio2026}.

\subsection*{Notation}
\begin{itemize}
\item Dirac's notation is used to denote states.
\item Repeated indices understand a summation (unless otherwise stated).
\item Latin letters denote space indices, while greek letters denote spacetime indices. A label $0$ indicates the time component of a four-vector.
\item A spatial three-vector is written with an arrow on top, e.g.~$\vec{x}$ for the spatial coordinates.
\item $\epsilon_{ijk}$ is the totally antisymmetric Levi-Civita symbol (with $\epsilon_{123}=1$).
\item The convention for the spacetime metric is diag$(1,-1,-1,-1)$.  Its components $\eta_{\mu\nu}$ are used as usual to lower and raise the spacetime indices.
\item A dot denotes a time derivative.
\item We use natural units where the speed of light, the reduced Planck constant and the Boltzmann constant are all equal to one.
\item $O^\dagger$: Hermitian conjugate of a generic linear operator $O$.
\item $A^*$: complex conjugate of a  generic  matrix $A$.
\item $A^T$: transpose of $A$.
%\item $A^{\dag}$: Hermitian conjugate of $A$.
\item $[O_1,O_2]$: commutator of two generic operators $O_1$ e $O_2$.
\item $\{O_1,O_2\}$: anticommutator of two generic operators $O_1$ e $O_2$.
\item $\theta$ is the Heaviside step function and 
$\delta$ is the Kronecker or Dirac delta function.
\end{itemize}

\section{General equilibrium density matrix }\label{density matrix}

The most general equilibrium density matrix $\rho_{\rm gen}$ is an exponential function of the Hamiltonian $H$ and the other conserved quantities, which commute with $H$ (See e.g.~\citet{Nair2005} and Sec.~2 of I1 for an introduction). So  in the relativistic case
  \be \rho_{\rm gen}= \frac1{Z}\exp\left(-\beta_\mu P^\mu-\frac12\beta_{ij}J_{ij} - \beta_a Q^a\right), \label{EqDen}\ee 
  where $P^0=H$, the $P^i$ are the components of the (linear) momentum, $J_k \equiv  \frac12 \epsilon_{ijk}J_{ij}$ (with $J_{ij}=-J_{ji}$) are the components of  the angular momentum, the $Q^a$ are the full set of charges, which generate a possible continuous internal (not necessarily Abelian) symmetry group $\mathcal{G}$, if any, and the $\beta_\mu$, $\beta_{ij}$ and $\beta_a$ are real constants. Since $J_{ij}=-J_{ji}$ we impose $\beta_{ij}=-\beta_{ji}$ without loss of generality.   The quantity $Z$ in~(\ref{EqDen}) is the  partition function:
  \be Z = \Tr\exp\left(-\beta_\mu P^\mu-\frac12\beta_{ij}J_{ij} - \beta_a Q^a\right). \label{ZTr}\ee 
  
  \subsection{Transition to the rest frame and conversion formul\ae~}\label{Rest frame}
   Let us now write $\beta_{ij} J_{ij} = 2\vec\tau \cdot \vec J$, where $\tau_k\equiv \frac12 \epsilon_{ijk}\beta_{ij}$  and $\vec\tau \cdot \vec J\equiv \tau_k J_k$, so that
  \be \rho_{\rm gen}= \frac1{Z}\exp\left(-\beta_0 H-\vec\beta\cdot \vec P-\vec\tau \cdot \vec J - \beta_a Q^a\right), \label{EqDen2}\ee
  with $\vec\beta\cdot \vec P \equiv \beta_i P^i$. Sometimes $\vec\tau$  is named thermal vorticity.
  Also, we can decompose $\vec\beta$ as $\vec\beta = \vec\beta^{\parallel}+\vec\beta^{\bot}$, where $\vec\beta^{\parallel}$ and $\vec\beta^{\bot}$ are, respectively, parallel and orthogonal to $\vec\tau$. The operator $\vec\beta^{\parallel}\cdot \vec P$ commutes with $\beta_0 H,~\vec\beta^{\bot}\cdot \vec P,~\vec\tau \cdot \vec J$ and $\beta_a Q^a$, but $\vec\beta^{\bot}\cdot \vec P$ only commutes with $\beta_0 H,~\vec\beta^{\parallel}\cdot \vec P$ and $\beta_a Q^a$, not with $\vec\tau \cdot \vec J$. 
  
  The technical complications due to $[\vec\beta^{\bot}\cdot \vec P,\vec\tau \cdot \vec J]\neq 0$ can be avoided by considering a space translation of the reference frame:
  \be U(\vec a) \rho_{\rm gen} U(\vec a)^\dagger = \frac1{Z}\exp\left(-\beta_0 H-\vec\beta\cdot \vec P-\frac12\beta_{ij}(J_{ij}+a_iP_j-a_jP_i) - \beta_a Q^a\right),\ee
  where $U(\vec a) = \exp(-i a_jP^j)$ is the unitary operator representing the space translation associated with $\vec a$ and the well-known commutation rules between the linear and angular momentum and the Hamiltonian of a relativistic theory,
  \be   i[P^j, J_{ik}] = \delta_{ji} P^k-\delta_{jk} P^i,\label{The Poincar algebraF}  
\ee
have been used. From $\beta_{ij}=\epsilon_{ijk}\tau_k$ one obtains
   \be  U(\vec a) \rho_{\rm gen} U(\vec a)^\dagger= \frac1{Z}\exp\left(-\beta_0 H-(\vec\beta-\vec\tau\times\vec a)\cdot \vec P-\vec\tau \cdot \vec J - \beta_a Q^a\right)\ee
   and we can find $\vec a$ such that $\vec\tau\times\vec a=-\vec\beta^\bot$, so
   \be  U(\vec a) \rho_{\rm gen} U(\vec a)^\dagger= \frac1{Z}\exp\left(-\beta_0 H-\vec\beta^\parallel\cdot \vec P-\vec\tau \cdot \vec J - \beta_a Q^a\right). \ee
   In this new reference frame the density matrix is a function of $\beta_0 H$, $\vec\beta^\parallel\cdot \vec P$, $\vec\tau \cdot \vec J$ and $\beta_a Q^a$, which commute with each other.

   Now, the convergence of the trace
    in~(\ref{ZTr})
   requires\footnote{In Sec.~2 of  I1 the simple case $\vec\tau=0$ was assumed, but one can consider the terms in the trace in~(\ref{ZTr}) with vanishing eigenvalue of $\vec\tau\cdot \vec J$ and so $\beta_0>0$ and $|\beta^\parallel|<\beta_0$ remain necessary for that convergence.} $\beta_0>0$ and $|\vec\beta^\parallel|<\beta_0$ (I1), meaning that $\{\beta_0, \vec\beta^\parallel\}$ form a time-like vector. So  one can now consider a Lorentz transformation $\Lambda$, implemented by a unitary operator $U(\Lambda)$, such that $U(\Lambda)\, \vec\tau \cdot \vec J \, U(\Lambda)^\dagger=\vec\tau \cdot \vec J$ and 
    \be U(\Lambda)U(\vec a) \rho_{\rm gen} U(\vec a)^\dagger U(\Lambda)^\dagger = \frac{e^{-\beta H-\vec\tau  \cdot \vec J - \beta_a Q^a}}{Z} \equiv \rho, \label{rhoRest0}
  \ee
  where $\beta\equiv1/T \equiv \sqrt{\beta_0^2-|\vec\beta^\parallel|^2} >0$ is the inverse temperature. To obtain~(\ref{rhoRest0}) note that  a Lorentz transformation that brings the time-like vector $\{\beta_0, \vec\beta^\parallel\}$ into $\{\beta, 0\}$ can involve only the component of the boost operator along $\vec\tau$, which commutes with $\vec\tau\cdot\vec J$. 
  
  It is convenient to express $\rho$ in the following form  (\cite{Landau1969},~\cite{Becattini2012},~\cite{Zubarev1979} and~\cite{Weert1982})
   \be \rho = \frac{e^{-\beta (H-\vec\Omega  \cdot \vec J - \mu_a Q^a)}}{Z}, \label{rhoRest}
  \ee
  where $\mu_a\equiv - \beta_a/\beta$ is the chemical potential associated with $Q^a$ and $\vec\Omega\equiv -\vec \tau/\beta$.   The operator
$\rho$ represents here the density matrix in the frame  where\footnote{For a generic operator $\mathcal{F}$ the ensemble average is 
$\langle \mathcal{F}\rangle =  \Tr(\rho \mathcal{F})$.
} $\langle \vec P\rangle = \Tr(\rho\vec P) =  0$ (see Sec.~2 of I1), which we adopt from now on. One can obtain the corresponding formul\ae~for the density matrix in a generic inertial  frame, $\rho_{\rm gen}$, by inverting~(\ref{rhoRest0}):
\be \rho_{\rm gen}  =U(\vec a)^\dagger U(\Lambda)^\dagger\rho U(\Lambda)U(\vec a). \label{rhoRestInv}
  \ee
  
  \subsection{Ensemble averages} \label{Ensemble averages}
  From~(\ref{rhoRest0}) one obtains the following expressions for the (ensemble) average energy, angular momentum and charge: respectively,
  \bea \langle H\rangle &=&  \Tr\left(\rho H\right)=\left.-\partial_\beta \log Z\right|_{\vec\tau, \beta_a}, \label{avHg}\\
   \langle J_k\rangle &=&  \Tr\left(\rho J_k\right)=\left.-\partial_{\tau_k} \log Z\right|_{\beta, \beta_a},\label{avJg} \\
   \langle Q^a \rangle &=& \Tr\left(\rho Q^a\right)=\left.-\partial_{\beta_a} \log Z\right|_{\beta, \vec\tau}.  \label{avQg} \eea
The derivatives on the right-hand sides are, respectively, taken keeping $\{\vec\tau, \beta_a\}$, $\{\beta, \beta_a\}$ and $\{\beta, \vec\tau\}$ constant as indicated by $\left. ... \right|_{\vec\tau, \beta_a}$, $\left. ... \right|_{\beta, \beta_a}$ and $\left. ... \right|_{\beta, \vec\tau}$. 

Rotational symmetry tells us that the average angular momentum $\langle \vec J\,\rangle = \Tr(\rho\vec J\, )$ is parallel to $\vec\Omega$ (or equivalently $\vec\tau$): 
\be \langle\vec J\, \rangle = \alpha_J \vec\Omega, \ee 
where $\alpha_J$ is a function of the rotationally invariant quantities $\beta$, $\mu_a$ and $\Omega\equiv|\vec \Omega|$.
  
  %read printed to here
   
  \section{Free fields}\label{Free fields}
  Let us start with the simplest case of free particles, which are described by free fields. General interacting theories will be investigated in Sec.~\ref{Interacting theories and general path-integral formulas}.
 
  \subsection{Scalars}\label{Free Scalars}

  First, let us consider a generic number of free scalar fields,  $\varphi_s$ (in this chapter all $\varphi_s$ are taken to be real without loss of generality).  Here the Lagrangian is 
 \be \mathscr{L}= \frac12\partial_\mu \varphi \partial^\mu\varphi - \frac12 \varphi M^2 \varphi, 
  \label{freeLag}\ee
where 
$M^2$ is the scalar squared-mass matrix. 
Note that $\varphi$ is a field array with components $\varphi_s$ and the transpose operation is understood.

The group $\mathcal{G}$ acts on $\varphi$ as follows: 
\be \varphi \to \exp(i\alpha_a\theta^a) \varphi \ee
for some real parameters $\alpha_a$, 
where the $\theta^a$ are the generators of $\mathcal{G}$ in the scalar representation. Since the $\varphi_s$ are taken to be real, $\theta^a$ are purely imaginary and antisymmetric: this ensures that the group has unitary elements and preserves the reality of the scalar fields. The invariance of the mass terms in~(\ref{freeLag}) implies $[\theta^a,M^2]=0$, which in turn tells us (by Schur's Lemma) that $M^2$ can be taken to be diagonal with blocks proportional to the identity matrix;
 the different blocks correspond to irreducible representations of $\mathcal{G}$. 
 This allows us to consider, at least for free fields, the various irreducible representations separately as we do from now on in this Sec.~\ref{Free Scalars}. All fields belonging to the same irreducible representation have the same mass, which in the following is denoted\footnote{The letter $m$ is not used for the mass here because it is used for the angular-momentum quantum number, see below.} $\mu$. The generators of $\mathcal{G}$ in the given irreducible representation are denoted $R^a$.

The corresponding field operator $\Phi$ is the most general solution of the Klein-Gordon equation
\be (\Box +\mu^2)\Phi = 0, \label{KG}\ee 
with $\Box\equiv\partial_\mu\partial^\mu$, satisfying the canonical commutators
\be [\Phi_s(t,\vec x), P_{\varphi\, s'}(t,\vec y)] = i\delta_{ss'} \delta(\vec x-\vec y), \quad  [\Phi(t,\vec x), \Phi(t,\vec y)] = 0, \quad  [P_{\varphi}(t,\vec x), P_{\varphi}(t,\vec y)] = 0, \label{CanComm}\ee
where $P_{\varphi} = \dot\Phi$ is the conjugate momentum operator and a dot denotes a time derivative. As well known, Eqs.~(\ref{KG}) and~(\ref{CanComm}) imply that $\Phi$ can be written in terms of creation and annihilation operators. Here, however, it is convenient to use the unusual cylindrical coordinates 
\be x^1 = r\cos\phi, \qquad x^2 = r\sin\phi, \qquad x^3 = z\label{CyCoo} \ee
instead of the Cartesian coordinates, $\{x^1, x^2, x^3\}$, with the $z$ axis chosen to be parallel to $\vec\Omega$. 
In cylindrical coordinates~\citep{Vilenkin1980}
\be \Phi_s(t,\vec x) =\sum_{m=-\infty}^{+\infty}\int_\mu^\infty d\omega\int_{-p_0}^{p_0}dp \left[a_{\omega p m s} \varphi_{\omega p m}(\vec x) e^{-i\omega t} +\mbox{h.c.}\right], \label{PhiDec} \ee
where the $a_{\omega p m s}$ are the annihilation operators of particles with energy $\omega$, linear and angular  momentum along the third axis $p$ and $m$, respectively, and species $s$, 
\be [a_{\omega p m s}, a_{\omega' p' m' s'}^\dagger] = \delta_{ss'} \delta_{mm'}\delta(\omega - \omega')\delta(p-p') \label{Commaad}\ee
and
\be \varphi_{\omega p m}(\vec x) e^{-i\omega t} = \frac{J_m(\alpha r)}{2^{3/2}\pi} e^{ipz+im\phi-i\omega t} \label{WFcyl}\ee 
are the corresponding wave functions, with $\alpha \equiv \sqrt{p_0^2-p^2}$, 
$p_0\equiv \sqrt{\omega^2 -\mu^2}$ and the $J_m$ being the cylindrical Bessel functions. A way to show that~(\ref{PhiDec}) is the  correct expression is to note that a generic linear combination of the wave functions in~(\ref{WFcyl}) is the general regular solution of the Klein-Gordon equation~(\ref{KG}) and the canonical commutators in~(\ref{CanComm}) imply~(\ref{Commaad}) with the normalization given in~(\ref{WFcyl}). To see that the canonical commutators in~(\ref{CanComm}) imply~(\ref{Commaad}) note that the canonical commutators imply the commutation relations of the annihilation and creation operators in Cartesian coordinates and the annihilation operators in cylindrical coordinates can be expressed as unitary linear  combinations of those in Cartesian coordinates.

Eqs.~(\ref{avHg}),~(\ref{avJg}) and~(\ref{avQg}) are the general formul\ae~to compute the  averages of the observables $H$, $J_k$ and $Q^a$ for arbitrary equilibrium density matrices. In this section we provide a general method to  explicitly determine the relevant averages in the case of free scalar fields. In more physical terms the case we are going to consider in this section is that of systems with negligibly small interactions. 

To facilitate the computation of averages let us perform a change of basis in the space of particle states as follows. 
Note that the action of $\mathcal{G}$ on $\Phi(x)$ is
\be \exp(i\alpha_aQ^a)\Phi(x)\exp(-i\alpha_aQ^a) = \exp(i\alpha_aR^a)\Phi(x), \label{GonPhi} \ee
which corresponds to the following action on the annihilation and creation operators,
\bea  \exp(i\alpha_aQ^a)a_{\omega p m s}\exp(-i\alpha_aQ^a) = \exp(i\alpha_aR^a)_{ss'}a_{\omega p m s'}, \\ 
\exp(i\alpha_aQ^a)a^\dagger_{\omega p m s}\exp(-i\alpha_aQ^a) = \exp(i\alpha_aR^a)_{ss'}a^\dagger_{\omega p m s'} . \label{Gonad} \eea
This implies the following action of $\mathcal{G}$  on one-particle states $|\omega,p,m,s\rangle$, namely states that are obtained by applying $a_{\omega p m s}^\dagger$ on the vacuum:
  \be  \exp(i\alpha_aQ^a) |\omega,p,m,s\rangle =  \exp(i\alpha_aR^a)_{ss'} |\omega,p,m,s'\rangle, \label{Gon1P} \ee
  where the invariance of the vacuum under $\mathcal{G}$ was used; the $|\omega,p,m,s\rangle$ are one-particle states with energy $\omega$, linear and angular  momentum along the third axis $p$ and $m$, respectively, and species $s$.
The expression above implies, given the generality of the $\alpha_a$,
  \be \mu_a Q^a|\omega,p,m,s\rangle =  (\mu_a R^a)_{ss'} |\omega,p,m,s'\rangle. \ee
 By performing a $\mu_a$-dependent unitary transformation of these states,
  \be |\omega,p,m;d\rangle\equiv U_{ds} |\omega,p,m,s\rangle, \ee 
  with the $U_{ds}$ satisfying $U_{ds}U^*_{d's}= \delta_{dd'}$, it is possible to diagonalize the matrix $\mu_aR^a$:
  \be U\mu_a R^aU^\dagger = \mathcal{M}, \label{fromRtoD} \ee 
  where $U$ is the matrix with elements $U_{ds}$ and $\mathcal{M}$ is a (generically $\mu_a$-dependent) diagonal real matrix. Therefore, in the new   basis 
    \be \mu_a Q^a|\omega,p,m;d\rangle =  \mathcal{M}_d |\omega,p,m; d\rangle, \label{muQM}\ee
    where the $\mathcal{M}_d$ are the diagonal elements of $\mathcal{M}$ and in~(\ref{muQM}) there is no sum over the index $d$. The $\mathcal{M}_d$ encode the effect of the chemical potentials.

  This basis is convenient to compute $\rho$, $Z$ and the averages of the product of two annihilation and creation operators and of the observables $H$, $\vec J$ and $Q^a$. To this purpose, another convenient thing to do is to discretize the variables  $\omega$ and $p$  such that integrals over these quantities become sums: this can be done by dividing the ranges of $\omega$ and $p$ in small discrete steps of size $\Delta\omega$ and  $\Delta p$. Such discretization can be obtained by putting the system in a finite volume, for example, of cylindrical shape (with axis in the direction of $\vec\Omega$) and imposing the periodicity condition $e^{ip(z+L)}=e^{ipz}$ and the continuity condition\footnote{This condition corresponds to imposing the vanishing of the field at the radial boundary of the system, which is nothing but the statement that the system has a finite radial size. In the presence of rotation, this can be interpreted as the effect of a centripetal force that balances the centrifugal one.} $J_m(\alpha R)=0$, where $R$ and $L$ are the radius and height of the cylinder. As usual $e^{ip(z+L)}=e^{ipz}$ implies that one can write $p=2\pi j/L$, where $j$ is a generic integer. To find the discrete values of $\omega$ note that $J_m(\alpha R) = 0$
   implies that $\alpha$ is quantized: for each value of $m$, the quantity $\alpha$  can only assume values 
  \be \alpha_{m,n} \equiv \frac{j_{m,n}}{R}, \label{alphadef}\ee
   where $j_{m,n}$ is the $n$th positive zero of the function $J_m$. So  for each $m$ and $p$ the energy can only assume the discrete values 
\be \omega_{m,n}(p) \equiv \sqrt{\mu^2+ \alpha_{m,n}^2 + p^2},\qquad n = 1,2,3,...\, . \label{omegamn} \ee 
In the discretized version, for any integrand $(...)$ one should actually read
\be \SumInt_q (...)\equiv \sum_{m=-\infty}^{+\infty}\int_\mu^\infty d\omega\int_{-p_0}^{p_0}dp \left(...\right) = \sum_{m=-\infty}^{+\infty}\sum_{j=-\infty}^{+\infty}\frac{2\pi}{L} \sum_{n=1}^\infty  \Delta \omega_{m,n}(p) \left(...\right)^{\omega\to\omega_{m,n}(p)}_{p\to 2\pi j/L} , \label{DicMes} \ee
where 
\be \Delta \omega_{m,n}(p) \equiv \omega_{m,n+1}(p) - \omega_{m,n}(p), \ee 
and we have also introduced a shorthand notation, with $q$ representing the set of variables $\{\omega, p, m\}$. As explicitly indicated in the right-hand side of~(\ref{DicMes}), the continuous variables $\omega$ and $p$ should be substituted by the respective discretized versions, $\omega_{m,n}(p)$ and\footnote{Since  $p=2\pi j/L$, where $j$ is an integer, as usual the integral measure can be obtained from $1/L= \Delta n/L =  dp/2\pi$.} $2\pi j/L$,   in the generic integrand  $(...)$.
  
After discretizing, one can introduce the rescaled annihilation operators $\alpha_{\omega p m, d} \equiv \sqrt{\Delta\omega\Delta p} \, a_{\omega p m, d}$ (with $a_{\omega p m, d}\equiv U_{ds}^*a_{\omega p m s}$), which satisfy
    \be [\alpha_{\omega p m, d}, \alpha_{\omega' p' m', d'}^\dagger] = \delta_{dd'}\delta_{mm'} \delta_{\omega \omega'}\delta_{pp'}. \label{CommaadD}\ee

    The density matrix in~(\ref{rhoRest}) can then be expressed in terms of the number operators 
    \be N_{\omega p m d} \equiv \alpha^\dagger_{\omega p m, d}\alpha_{\omega p m, d}\ee 
    as follows:
    \be \rho=\frac1{Z} \exp\left(-\beta \sum_{\omega p m d}\left( \omega  -m\Omega   - \mathcal{M}_d  \right)N_{\omega p m d}\right) . \ee 
  This is nothing but the density matrix with zero thermal vorticity and chemical potentials, but with energies $\omega$ replaced by $\omega - m\Omega - \mathcal{M}_d$. As a result the partition function is, cf.~Sec.~3.1 of I1,
    \be Z = \prod_{\omega p m d} \frac1{1-e^{-\beta\left(\omega-m\Omega - \mathcal{M}_d\right)}}. \label{Zpart}\ee
    Then, using  
\be \log Z =  - \sum_{\omega p m d} \log \left(1- e^{-\beta(\omega - m\Omega   - \mathcal{M}_d)}  \right), \label{logZ}\ee 
one finds
\bea \langle \alpha_{\omega p m, d}^\dagger \alpha_{\omega' p' m', d'}\rangle &=& f_B(\omega - m\Omega - \mathcal{M}_d)\delta_{dd'} \delta_{mm'} \delta_{\omega \omega'}\delta_{pp'}, \label{dalpha}\\ 
\langle \alpha_{\omega p m, d} \alpha^\dagger_{\omega' p' m', d'}\rangle &=& (1+f_B(\omega - m\Omega - \mathcal{M}_d)) \delta_{dd'} \delta_{mm'}\delta_{\omega \omega'}\delta_{pp'}, \label{alphad}\eea
where
\be f_B(x) \equiv \frac1{e^{\beta x}-1} \label{fBdef}\ee
is the Bose-Einstein distribution. Setting $d=d'$, $m=m'$, $\omega=\omega'$ and $p=p'$ in~(\ref{dalpha}) one finds the average numbers of particles. Moreover,  as usual, 
\be \quad \langle \alpha_{\omega p m, d} \alpha_{\omega' p' m', d'}\rangle = \langle \alpha_{\omega p m, d}^\dagger \alpha_{\omega' p' m', d'}^\dagger\rangle = 0. \ee
Let us recall that in~(\ref{Zpart}) a single irreducible representation of $\mathcal{G}$ is considered: to obtain the partition function for all  irreducible representations one can simply take the product of all partition functions of single irreducible representations.

Moreover, using~(\ref{logZ}) and the general expressions~(\ref{avHg}),~(\ref{avJg}) and~(\ref{avQg}), the average values of $H$, $\vec J$  and $Q^a$ read 
\be \langle H\rangle =  
\sum_{\omega p m d} \omega f_B(\omega - m\Omega - \mathcal{M}_d), \qquad   \langle  J_z\rangle =  
\sum_{\omega p m d} m f_B(\omega - m\Omega - \mathcal{M}_d), \qquad \langle Q^a\rangle =  
\sum_{\omega p m d} \frac{\partial\mathcal{M}_d}{\partial\mu_a}  f_B(\omega - m\Omega - \mathcal{M}_d) \label{HJzQ}.\ee
In the limit $\Delta p\to0$ (which corresponds to $L\to\infty$) these expressions allow us to compute  the average values of the energy, angular momentum and charges per unit of length in the $z$ direction in terms of integrals rather than sums over $p$.

Let us note that we can also go back to the original basis by inverting~(\ref{fromRtoD}) and obtain, starting from~(\ref{dalpha})-(\ref{alphad}), 
  \bea \langle \alpha_{\omega p m s}^\dagger \alpha_{\omega' p' m' s'}\rangle &=& f_B(\omega - m\Omega - \mu_a R^a)_{ss'}\delta_{mm'} \delta_{\omega \omega'}\delta_{pp'},\label{Avada} \\ 
\langle \alpha_{\omega p m s} \alpha^\dagger_{\omega' p' m' s'}\rangle &=& (1+f_B(\omega - m\Omega - \mu_a R^a))_{s's}\delta_{mm'} \delta_{\omega \omega'}\delta_{pp'}, \label{Avaad}\eea
with $\alpha_{\omega p m s} \equiv U_{ds}\alpha_{\omega p m, d}$, and 
\be \quad \langle \alpha_{\omega p m s} \alpha_{\omega' p' m' s'}\rangle = \langle \alpha_{\omega p m s}^\dagger \alpha_{\omega' p' m' s'}^\dagger\rangle = 0. \label{Avaa}\ee 
The averages in the left-hand sides of~(\ref{Avada}) and~(\ref{Avaad})  were  expressed in terms of series by~\citet{Becattini2021}.
The quantity $f_B(\omega - m\Omega - \mu_a R^a)$ appearing in the expressions above is the function $f_B$ in~(\ref{fBdef}) computed in the matrix $\omega-m\Omega- \mu_a R^a$ (where an identity matrix multiplying $\omega-m\Omega$ is understood). In practice, since $\mu_a R^a$ is a Hermitian and thus diagonalizable matrix, to compute this quantity one can determine the projectors $\mathcal{P}_d$ associated with the eigenvalues $\mathcal{M}_d$ of  $\mu_a R^a$ and use the well-known spectral decomposition
\be f_B(\omega - m\Omega - \mu_a R^a) = \sum_d f_B(\omega - m\Omega - \mathcal{M}_d)\mathcal{P}_d, \label{SpDec}\ee 
where the projectors satisfy $\sum_d\mathcal{P}_d =1$, $\mathcal{P}_d\mathcal{P}_{d'} = \delta_{dd'}\mathcal{P}_{d'}$ and $\mu_a R^a = \sum_d\mathcal{M}_d \mathcal{P}_d$.  Both the eigenvalues $\mathcal{M}_d$ and the projectors $\mathcal{P}_d$ generically depend on the chemical potentials $\mu_a$.

To have a feeling of how the $\mathcal{M}_d$ and $\mathcal{P}_d$ may look like in some non-Abelian case, consider $\mathcal{G} =$ SO(3) and a triplet scalar representation, so
\be R^1=i  
\left(\begin{array}{ccc}
 0 & 0 & 0 \\
 0 & 0 & -1 \\
 0 & 1 & 0 \\
\end{array}\right), \quad  R^2=i \left(
\begin{array}{ccc}
 0 & 0 & 1 \\
 0 & 0 & 0 \\
 -1 & 0 & 0 \\
\end{array}
\right), \quad R^3=i \left(
\begin{array}{ccc}
 0 & -1 & 0 \\
 1 & 0 & 0 \\
 0 & 0 & 0 \\
\end{array}
\right). \label{R123}
\ee
In this case the eigenvalues of $\mu_aR^a$ are $\mathcal{M}_0 =0$ and $\mathcal{M}_{\pm} = \pm \sqrt{\mu_a\mu_a}$ and the corresponding projectors read
\be \mathcal{P}_0=\left(
\begin{array}{ccc}
 \frac{\mu_1^2}{\mu_a\mu_a}& \frac{\mu_1 \mu_2}{\mu_a\mu_a} & \frac{\mu_1 \mu_3}{\mu_a\mu_a} \\
 \frac{\mu_1 \mu_2}{\mu_a\mu_a} & \frac{\mu_2^2}{\mu_a\mu_a} & \frac{\mu_2 \mu_3}{\mu_a\mu_a} \\
 \frac{\mu_1 \mu_3}{\mu_a\mu_a} & \frac{\mu_2 \mu_3}{\mu_a\mu_a} & \frac{\mu_3^2}{\mu_a\mu_a} \\
\end{array}
\right), \quad \mathcal{P}_\pm=\left(
\begin{array}{ccc}
 \frac{\mu_2^2+\mu_3^2}{2 \mu_a\mu_a} & -\frac{\mu_1 \mu_2+i\mu_3\mathcal{M}_\pm }{2 \mu_a\mu_a} & \frac{i\mu_2\mathcal{M}_\pm-\mu_1 \mu_3}{2 \mu_a\mu_a} \\
 \frac{i\mu_3\mathcal{M}_\pm-\mu_1 \mu_2}{2 \mu_a\mu_a} & \frac{\mu_1^2+\mu_3^2}{2 \mu_a\mu_a} & -\frac{i\mu_1\mathcal{M}_\pm +\mu_2 \mu_3}{2 \mu_a\mu_a} \\
 -\frac{i\mu_2\mathcal{M}_\pm+\mu_1 \mu_3}{2 \mu_a\mu_a} & \frac{i\mu_1\mathcal{M}_\pm-\mu_2 \mu_3}{2 \mu_a\mu_a} & \frac{\mu_1^2+\mu_2^2}{2 \mu_a\mu_a} \\
\end{array}
\right).\ee

It is important to note that the convergence of the averages requires $\Omega$ to be less than a critical value. To understand this point let us consider $\log Z$ in~(\ref{logZ}), which can be used to extract the averages as explained above. 
Separately considering the sum over $m$, the convergence of the series at $m\to +\infty$ implies 
\be \Omega <  \lim_{m\to +\infty}\frac1{m}\min_{\omega\, d} ( \omega - \mathcal{M}_d). \ee 
Using now the expression for the energies in~(\ref{omegamn}), this condition reads
\be \Omega <  \lim_{m\to +\infty}\frac1{m}\min_{n\, p\, d} ( \omega_{m,n}(p) - \mathcal{M}_d) = \lim_{m\to +\infty} \frac{j_{m,1}}{m R} = \frac1{R},  \label{ConvCond} \ee 
where we have used that $j_{m,1}/m$ converges to 1 as\footnote{Cf.~\href{https://dlmf.nist.gov/10.21}{https://dlmf.nist.gov/10.21}.} $m\to+\infty$.
From~(\ref{HJzQ}) one can clearly see that this condition is implied by the physical requirement that the average energy, angular momentum and charges should be finite. In Sec.~\ref{Interacting theories and general path-integral formulas} we will see that $\vec \Omega$ can be identified with the angular-velocity vector of the rotating plasma, so $R\Omega<1$ is nothing but the physical requirement  that the plasma must not reach the speed of light.

Some applications and consequences of the results  presented in this Sec.~\ref{Free Scalars} are discussed in~\citet{Salvio2025b}, where the results were derived.

\subsection{Vectors}\label{Free Vectors}

It is easy to extend the result of Sec.~\ref{Free Scalars} regarding the  averages for free spin-0 particles to free spin-1 particles, which are described by vector fields. One can write spin-1 one-particle states in a basis of eigenstates of $H$, $P^z$, $J_z$ and $\vec J\cdot \vec P/|\vec P|$, with the latter operator, the helicity, having two (three) eigenvalues for massless (massive) spin-1 particles;
   this corresponds to the respective two (three) degrees of freedom of a massless (massive) spin-1 particle. Adapting the discussion of Sec.~\ref{Free Scalars}, one then finds that the partition function of a free spin-1 particle is the same as  the partition function of a free spin-0 particle in the same representation of $\mathcal{G}$ raised to a power equal to the number of degrees of freedom.  Then the  averages of $H$, $\vec J$ and $Q^a$ for a free spin-1 particle are equal to  the corresponding quantities for a free spin-0 particle but multiplied by the number of degrees of freedom. 

Then the convergence of the averages for spin-1 particles, just like the convergence of the  averages for spin-0 particle,
requires the bound $\Omega<1/R$.

  \subsection{Fermions}\label{fermion Free fields}
  Let us now consider the case of spin-1/2 fermions. 
  We start with a generic number of free Dirac  fields,  $\psi_s$, with Dirac masses. Later on also Weyl fields as well as Majorana masses will be studied. 
  The Lagrangian is here given by 
 \be \mathscr{L}= \bar \psi (i\slashed{\partial}-\mu_F) \psi,
  \label{freeLagf}\ee
where 
$\mu_F$ is the (Dirac) mass matrix related to the fermion squared mass matrix $M_F^2$ through $M_F^2\equiv \mu_F\mu_F^\dagger$. 
Like for the scalar fields a vector notation is used, $\psi$ is a field array with components $\psi_s$ and, as usual, $\bar\psi \equiv \psi^\dagger \gamma^0$ and $\slashed{\partial}\equiv \gamma^\mu \partial_\mu$, where  $\gamma^\mu$ are the Dirac matrices, which satisfy $\{\gamma^\mu,\gamma^\nu\}=2\eta^{\mu\nu}$. 
   
   The group $\mathcal{G}$ acts on $\psi$ as follows: 
\be \psi \to \exp(i\alpha_a t^a) \psi \ee
for some real parameters $\alpha_a$, 
where the $t^a$ are the generators of $\mathcal{G}$  in the fermion representation. The $t^a$ are Hermitian matrices.  The invariance of the mass terms in~(\ref{freeLagf}) implies $[t^a,\mu_F]=0$, which in turn tells us (by Schur's Lemma) that $\mu_F$ can be taken to be diagonal with blocks proportional to the identity matrix;
 the different blocks correspond to irreducible fermion representations of $\mathcal{G}$.  
 This allows us to consider, at least for free fields, the various irreducible representations separately as we do from now on in this Sec.~\ref{fermion Free fields}. All fields belonging to the same irreducible representation have  the same mass, which in the following is denoted  $\mu$ like for scalars. 
 The generators of $\mathcal{G}$ in the given irreducible representation are denoted ${\cal R}^a$.
 
 The corresponding field operator $\Psi$ is the most general solution of the Dirac equation
\be (i\slashed{\partial}-\mu)\Psi = 0 \label{DEq}\ee 
satisfying the canonical anticommutators
\be \{\Psi_\alpha(t,\vec x), \Psi_\beta(t,\vec y)\} =0, \quad \{\Psi_\alpha(t,\vec x), \Psi_\beta^\dagger(t,\vec y)\} =\delta_{\alpha\beta} \delta(\vec x-\vec y). \label{CanAComm}\ee
 
Let us take now the cylindrical coordinates~(\ref{CyCoo}). One can work in the basis of eigenstates of the commuting operators $H$, $P^z$, $J_z$ and $\vec J\cdot \vec P/|\vec P|$. Let  us call $q$ the respective set of eigenvalues, which for the  fermion in question are  $\omega$, $p$, $m+1/2$ (with $m$ being  a generic integer) and $\sigma =\pm 1/2$.  Note that when $\mu=0$ one can consider just spinors that are eigenstates of chirality (twice the helicity, i.e.~$\gamma_5$ in the choice of~\citet{BjorkenDrell2005}) with a definite eigenvalue.
In general, one can write 
 \be \Psi_s(x) = \SumInt_q (\mathcal{U}_q(x) c_{qs}+ \mathcal{V}_q(x) d^\dagger_{qs}), \label{FreeFexp}\ee 
 where the integro-sum over $q$ is now defined, for any integrand $(...)$, by
\be \SumInt_q (...)\equiv \sum_{m=-\infty}^{+\infty}\sum_{\sigma=\pm1/2}\int_\mu^\infty d\omega\int_{-p_0}^{p_0}dp \left(...\right),
 \ee 
 $p_0\equiv \sqrt{\omega^2 -\mu^2}$, 
 the $\mathcal{U}_q$ and $\mathcal{V}_q$ are the complete set of solutions of~(\ref{DEq}) in this basis for particles and antiparticles, respectively, 
 and  the $c_{qs}$ and $d_{qs}$ are the corresponding annihilation operators for the fermion and antifermion, respectively, of species $s$. They satisfy
  \bea &&\{c_{q s},c_{q' s'}^\dagger\} = \{d_{q s},d_{q' s'}^\dagger\} =\delta(q-q')\delta_{ss'}, \label{antNot}\\ 
&&\{c_{q s},c_{q' s'}\}=\{d_{q s},d_{q' s'}\}=\{c_{q s},d_{q' s'}\}=\{c_{q s},d_{q' s'}^\dagger\}=0,\label{ant} \eea 
 where 
 \be \delta(q-q') \equiv  \delta_{mm'}\delta_{\sigma\sigma'}\delta(\omega - \omega')\delta(p-p'). \ee 
 Since the $\mathcal{U}_q$ and $\mathcal{V}_q$ form a  complete set of eigenfunctions of Hermitian operators (corresponding to the Hamiltonian, the linear and angular momentum along the third axis and the helicity)
  they can be normalized in a way that 
 \be \int d^3 x \, \mathcal{U}^\dagger_{q'}(x)\mathcal{U}_{q}(x) = \delta(q'-q)  = \int d^3 x \, \mathcal{V}^\dagger_{q'}(x)\mathcal{V}_{q}(x), \qquad  \int d^3 x \, \mathcal{V}^\dagger_{q'}(x)\mathcal{U}_{q}(x) =0  \label{OrtoF}\ee 
 and also 
 \be \SumInt_q( \mathcal{U}_q(t,\vec x)\mathcal{U}_q^\dagger(t,\vec y) +  \mathcal{V}_q(t,\vec x)\mathcal{V}_q^\dagger(t,\vec y)) = \delta(\vec x-\vec y). \label{completeF}\ee 
Since $^\dagger$ represents the conjugate transpose, the quantity in~(\ref{completeF}) is a matrix in spinor space. Using~(\ref{antNot}),~(\ref{ant}) and~(\ref{completeF}), one can check the anticommutation relations in~(\ref{CanAComm}). 
 The $\mathcal{U}_q$ and $\mathcal{V}_q$ contain the cylindrical Bessel functions. A way to see this is to note that $\Psi$, like the free scalar operator $\Phi$, satisfies the Klein-Gordon equation.

 The explicit form of the $\mathcal{U}_q$ and $\mathcal{V}_q$ depend on the basis choice for the $\gamma^\mu$. The explicit expression of $\mathcal{U}_q$ and $\mathcal{V}_q$ in the choice, for example, of~\citet{BjorkenDrell2005} was obtained by\footnote{Here a different normalization is used to implement~(\ref{OrtoF}) and~(\ref{completeF}), so the solutions denoted $U_j$ and $V_j$  for Dirac fermions of~\citet{Ambrus2016} are related to $\mathcal{U}_q$ and $\mathcal{V}_q$ through $\mathcal{U}_q =\sqrt{\omega} U_j$ and $\mathcal{V}_q=\sqrt{\omega} V_j$.}~\citet{Ambrus2016} (see also the previous study by~\citet{Vilenkin1979} for the case of chiral spinors). For this choice $\mathcal{V}_q = i \gamma^2 \mathcal{U}_q^*$.

 So far, only Dirac masses have been considered. In the limit where some of these masses vanish one can consider chiral fermions with just one helicity state (either $\sigma=+1/2$ or $\sigma=-1/2$) as a particular case. In order to present an analysis that is as general as possible, let us now also include  Majorana masses in the analysis. These appear in well-motivated extensions of the Standard Model, such as those featuring the type-I see-saw mechanism. The most suitable formalism to include Majorana masses is that of Weyl spinors,  which feature the following Lagrangian density 
 \be \mathscr{L}= \bar
 \psi i \slashed{\partial}\psi+\frac12 \left(\psi \mu_F\psi +\mbox{h.c.}\right).
  \label{freeLagW}\ee
  In the Weyl formalism we adopt the following notation.
\begin{itemize}
\item $\psi$ and $\bar\psi$ are two-component spinors with components $\psi_\alpha$ and $\bar\psi^\alpha$, respectively, ($\bar\psi^\alpha$ is interpreted here as the Hermitian conjugate of $\psi_\alpha$)
and the transpose operation is understood. We also introduce 
$\psi^\alpha\equiv \psi_\beta \epsilon^{\beta\alpha}$, $\bar\psi_\alpha\equiv \epsilon_{\alpha\beta}\bar\psi^\beta$, 
where $\epsilon^{\alpha\beta}$ and $\epsilon_{\alpha\beta}$ are the antisymmetric symbols with $\epsilon^{12}=1$ and $\epsilon_{12}=-1$, such that $\epsilon^{\alpha\beta}\epsilon_{\beta\gamma} = \delta^\alpha_{~\gamma}$.
\item The kinetic term $\bar
 \psi i \slashed{\partial}\psi$ is now constructed with the $2\times 2$ matrices $\bar\sigma^\mu$ (defined by $\{\bar{\sigma}^{\mu}\}\equiv (1, -\vec{\sigma})$ and $\vec{\sigma}$ represents the three Pauli matrices) as 
 \be\bar
 \psi i \slashed{\partial}\psi \equiv \bar
 \psi i \bar\sigma^\mu \partial_\mu\psi \equiv\bar\psi^\alpha i\bar\sigma^{\mu~\beta}_{~\alpha} \partial_\mu\psi_\beta. \ee
Neglecting boundary terms
 \be \int d^4x   \, \bar
 \psi i \slashed{\partial}\psi = \frac12  \int d^4x   \, \bar
 \psi i \slashed{\partial}\psi +
 \psi i \slashed{\partial}\bar \psi ,  \ee 
 with 
 \be \psi i \slashed{\partial}\bar \psi \equiv \psi i \sigma^\mu \partial_\mu \bar\psi\equiv  \psi^\alpha i \sigma^{\mu~\beta}_{~\alpha}\partial_\mu \bar\psi_\beta \ee
 and $\{\sigma^{\mu}\}\equiv (1, \vec{\sigma})$. Here we have used  $\sigma_2 \left(\bar \sigma^\mu\right)^T \sigma_2 = \sigma^\mu$. 
 \item Finally, $\mu_F$
 is the fermion mass matrix, which can include both Dirac and Majorana masses, and 
 \be \psi \mu_F \psi \equiv  \psi_{\beta i}\epsilon^{\beta\alpha} \mu_F^{ij} \psi_{\alpha j}, \qquad (\psi \mu_F\psi)^\dagger = \bar\psi_{\alpha j} (\mu_F^{ij})^* \epsilon^{\beta\alpha}\bar\psi_{\beta i}\equiv \bar\psi \mu_F^\dagger \bar\psi. \label{GenMass} \ee
 \end{itemize}
 Note that we can put  $\mu_F$ in diagonal form through a unitary transformation\footnote{This is known as the complex Autonne-Takagi factorization, see e.g.~\citet{Youla1961}.} acting on $\psi$.  
 We then  work with a field basis where $\mu_F$ is diagonal. The absolute values of the diagonal elements of $\mu_F$ are the fermion  masses. 
 Also in this case, in a given  irreducible representation of $\mathcal{G}$ all particles have the same masses by Schur's Lemma because the Hamiltonian always commutes with the $Q^a$ (the $Q^a$ are assumed to be conserved).
  The mass and generators in the given irreducible representation are denoted again $\mu$ and ${\cal R}^a$, respectively.  From now on in this Sec.~\ref{fermion Free fields} the various irreducible representations  are considered separately also for Weyl fields. This formalism allows us to easily describe also Majorana fermions.
 
The field operator $\Psi$ for a 
massive\footnote{The massless case can be easily treated through Dirac fields by considering eigenspinors  of chirality and imposing the condition that particles equal antiparticles, $({\cal R}^a)^* =  - \mathcal{R}^a$.} 
Majorana fermion is a solution of
\be i \slashed{\partial}\Psi = -\mu \epsilon \bar\Psi \label{WeylEq}\ee
instead of~(\ref{DEq}), satisfying the  canonical anticommutators
\be \{\Psi_\alpha(t,\vec x), \Psi_\beta(t,\vec y)\} =0, \quad \{\Psi_\alpha(t,\vec x), \bar \Psi^\beta(t,\vec y)\} =\delta_\alpha^{~\beta} \delta(\vec x-\vec y). \label{CanACommW}\ee
 Here $\epsilon$ is the $2\times2$ antisymmetric matrix with $\epsilon_{12}=-1$. 
 When working with Weyl-spinor operators, $\bar\Psi^\alpha$ represents the Hermitian conjugate of $\Psi_\alpha$.  In the case of Majorana fermions, which are described here by Weyl spinors for convenience, the decomposition of a general $\Psi_s$ (with $s$ being the species index) in terms of annihilation and creation operators of particle states with definite values of  $H$, $P^z$, $J_z$ and $\vec J\cdot \vec P/|\vec P|$ reads
    \be \Psi_s(x) = \SumInt_q (X_q(x) a_{qs}+Y_q(x) a^\dagger_{qs}), \label{FreeWexp}\ee
    where the $X_q$ and $Y_q$ are eigenspinors of  $H=i\partial_t$, $P^z$, $J_z$ and $\vec J\cdot \vec P/|\vec P|$ with eigenvalues $\{\omega,p,m+1/2,\sigma\}$ and $\{-\omega,-p,-m-1/2,\sigma\}$, respectively.
     Eq.~(\ref{WeylEq}) then implies
    \be (\omega+2|\vec p|\sigma) X_q +\mu \epsilon \bar Y_q =  (-\omega+2|\vec p|\sigma) Y_q +\mu \epsilon \bar X_q = 0, \label{EOMXY}\ee 
    where a bar on top of bispinors  represents a complex conjugate.
    Combining these two equations and using $\sigma=\pm 1/2$ leads to the on-shell relation $\omega^2=\mu^2+\vec p^2$. Moreover, both these equations allow us to express $Y_q$ in terms of $X_q$ (and viceversa):
    \be Y_q = \frac{\omega+2|\vec p|\sigma}{\mu} \epsilon \bar X_q. \label{YfromX}\ee
     
     Note that the Weyl field operator $\Psi_s$ in~(\ref{FreeWexp}) features only one type of annihilation operators, $a_{qs}$,
     as appropriate for Majorana fermions. Being a fermion system
 \be    \{a_{q s},a_{q' s'}^\dagger\} = \delta(q-q')\delta_{ss'}, \qquad \{a_{q s},a_{q' s'}\}  = 0 . \label{antNotW} \ee
Note also that $\mathcal{\bar  R}^a = \mathcal{R}^a$ because the antiparticle equals the particle.

One can easily show that the Majorana field $\Psi$ in~(\ref{FreeWexp}) satisfies the Klein-Gordon equation $\partial^2\Psi=-\mu^2\Psi$, which also follows from the on shell relation $\omega^2=\mu^2+\vec p^2$.
 As a result both $X_q$ and $Y_q$ are eigenfunctions of $\vec P^2$. Since the $X_q$  form a  complete set of eigenfunctions of Hermitian operators (corresponding to $\vec P^2$, 
$P^z$, $J_z$ and $\vec J\cdot \vec P/|\vec p|$)
  they can be normalized in a way that\footnote{With this normalization $2\pi \exp(i\omega t-i p z)X_q$ can be taken to be proportional to the quantitiy $\phi_j$ computed by~\citet{Ambrus2016}  with the proportionality factor given by $\sqrt{\mu c_x/2}$. Having determined $X_q$, the other  eigenspinor $Y_q$ is given by~(\ref{YfromX}). } 
 \be \int d^3 x \, \bar X_{q'}(x)X_{q}(x) = \frac{\mu c_x}{2\omega} \delta(q'-q),  \quad \mbox{with} \quad c_x \equiv \sqrt{\frac{\omega-2|\vec p|\sigma}{\omega+2|\vec p|\sigma}},  \label{OrtoFX}\ee 
 which corresponds to the completeness relation
 \be \SumInt_q X_{q\alpha}(t,\vec x)\bar X_q^\beta(t,\vec y) = \frac{\mu c_x}{2\omega}\delta_\alpha^{~\beta} \delta(\vec x-\vec y). \label{completeX}\ee 
 Using Eqs.~(\ref{EOMXY}) one finds the corresponding relations for the $Y_q$:
 \be \int d^3 x \, \bar Y_{q'}(x)Y_{q}(x) = \frac{\mu}{2\omega  c_x} \delta(q'-q), \label{OrtoFY}\ee
 and
 \be \SumInt_q Y_{q\alpha}(t,\vec x)\bar Y_q^{\beta}(t,\vec y) = \frac{\mu}{2\omega  c_x}\delta_\alpha^{~\beta} \delta(\vec x-\vec y). \ee 
 Using~(\ref{YfromX}),~(\ref{antNotW}) and~(\ref{completeX}) one can check the  canonical anticommutators in~(\ref{CanACommW}).

We show now a general method to  compute averages in the case of free fermion fields, i.e.~for systems involving fermions with negligibly small interactions.  
However,  in some cases one can take into account in-medium effects by substituting  the masses with effective masses and the chemical potentials with effective chemical potentials (examples were provided by~\cite{Liu2002} and~\cite{Fu2008}).  Sec.~\ref{Interacting theories and general path-integral formulas} will then furnish the methods to address theories with general interactions.

Like for scalars, to facilitate the computation of averages let us perform a change of basis in the space of particle states. This can  be done as follows. 
Note that $\mathcal{G}$ acts on $\Psi(x)$ as
\be \exp(i\alpha_aQ^a)\Psi(x)\exp(-i\alpha_aQ^a) = \exp(i\alpha_a{\cal R}^a)\Psi(x). \label{GonPsi} \ee
Let us first consider Dirac fermions, 
Eq.~(\ref{FreeFexp}). We will later clarify the modifications needed in the case of Majorana fermions, Eq.~(\ref{FreeWexp}).
Note that (\ref{GonPsi}) corresponds to the following action on the annihilation and creation operators,
\be  \exp(i\alpha_aQ^a)c_{qs}\exp(-i\alpha_aQ^a) = \exp(i\alpha_a{\cal R}^a)_{ss'}c_{q s'}, \qquad 
\exp(i\alpha_aQ^a)d_{q s}\exp(-i\alpha_aQ^a) = \exp(i\alpha_a\bar {\cal R}^a)_{ss'}d_{q s'}, \label{Gonacdf} \ee
where $\bar {\cal R}^a  = - ({\cal R}^a)^*$. This follows from the expressions of $c_{qs}$ and $d_{q s}$ in terms of $\Psi_s$, which can be obtained by inverting the decomposition in~(\ref{FreeFexp}) through the orthogonality conditions in~(\ref{OrtoF}).
The transformation rules in~(\ref{Gonacdf}) imply the following action of $\mathcal{G}$  on one-particle states, $|q,s\rangle\equiv c_{q s}^\dagger|0\rangle$ and $|\overline{q,s}\rangle\equiv d_{q s}^\dagger|0\rangle$  (these states have energy $\omega$, linear and angular  momentum along the third axis $p$ and $m+1/2$, respectively, helicity $\sigma$ and species $s$):
  \be  \exp(i\alpha_aQ^a) |q,s\rangle =  \exp(i\alpha_a\bar {\cal R}^a)_{ss'} |q,s'\rangle, \quad \exp(i\alpha_aQ^a) |\overline{q,s}\rangle =  \exp(i\alpha_a{\cal R}^a)_{ss'} |\overline{q,s'}\rangle,\label{Gon1Pf} \ee
  where  the invariance of the vacuum $|0\rangle$ under $\mathcal{G}$ was used.
The expressions above imply, among other things, 
  \be \mu_a Q^a|q,s\rangle =  (\mu_a \bar {\cal R}^a)_{ss'} |q,s'\rangle, \qquad \mu_a Q^a|\overline{q,s}\rangle =  (\mu_a {\cal R}^a)_{ss'} |\overline{q,s'}\rangle. \ee
    Now, by performing a $\mu_a$-dependent unitary transformation,
  \be |q;d\rangle\equiv W_{ds} |q,s\rangle, \quad |\overline{q;d}\rangle\equiv W^*_{ds} |\overline{q,s}\rangle \ee 
  (with the $W_{ds}$ satisfying $W_{ds}W^*_{d's}= \delta_{dd'}$) it is possible to diagonalize both $\mu_a\bar {\cal R}^a$ and $\mu_a{\cal R}^a$:
  \be W\mu_a \bar {\cal R}^aW^\dagger = \mathcal{M}^F, \quad W^*\mu_a {\cal R}^aW^T = -\mathcal{M}^F, \label{fromttoD} \ee 
  where $W$ is the matrix with elements $W_{ds}$ and $\mathcal{M}^F$ is a (generically $\mu_a$-dependent) diagonal real matrix. Therefore, in the new   basis 
    \be \mu_a Q^a|q;d\rangle =  \mathcal{M}^F_d |q; d\rangle, \quad \mu_a Q^a|\overline{q;d}\rangle =  -\mathcal{M}^F_d |\overline{q; d}\rangle, \label{muQMf}\ee
    where the $\mathcal{M}^F_d$ are the diagonal elements of $\mathcal{M}^F$ and in the right-hand side of~(\ref{muQMf}) there is no sum over the index $d$. The $\mathcal{M}^F_d$ encode the effect of the fermion chemical potentials.

    Like for scalars, we discretize the variables $\omega$ and $p$ and introduce the rescaled annihilation operators $\gamma_{q, d} \equiv \sqrt{\Delta\omega\Delta p} \, c_{q, d}$ and $\delta_{q, d} \equiv \sqrt{\Delta\omega\Delta p} \, d_{q, d}$ (with $c_{q,d}\equiv W_{ds}^*c_{q s}$ and $d_{q,d}\equiv W_{ds}d_{q s}$). The only non-vanishing anticommutators between rescaled annihilation and creation operators are
    \be \{\gamma_{q, d}, \gamma_{q', d'}^\dagger\} = \delta_{qq'}\delta_{dd'}, \quad \{\delta_{q, d}, \delta_{q', d'}^\dagger\} = \delta_{qq'}\delta_{dd'}, \label{ACommccdD}\ee
    with $\delta_{qq'}\equiv \delta_{mm'}\delta_{\sigma\sigma'} \delta_{\omega \omega'}\delta_{pp'}$. This discretization is useful to easily compute $\rho$, $Z$ and the  average of relevant quantities in full generality. The continuum limit (which corresponds to the large-volume limit in coordinate space) will be taken afterwards.

  The density matrix in~(\ref{rhoRest}) in the fermion case  can then be expressed in terms of the number operators for particles and antiparticles, respectively 
    \be N_{q d} \equiv \gamma^\dagger_{q, d}\gamma_{q, d}, \quad \bar N_{q d} \equiv \delta^\dagger_{q, d}\delta_{q, d}\ee 
    as follows:
    \be \rho=\frac1{Z} \exp\left(-\beta \sum_{q d}\left\{\left[ \omega  -(m+1/2)\Omega   - \mathcal{M}_d^F  \right]N_{q d}+\left[ \omega  -(m+1/2)\Omega   + \mathcal{M}_d^F  \right]\bar N_{q d}\right\} \right). \label{rhofref} \ee 
 The quantity in~(\ref{rhofref}) is nothing but the density matrix with zero thermal vorticity and chemical potentials, but with energies $\omega$ replaced by $\omega - (m+1/2)\Omega - \mathcal{M}_d^F$ for particles and by $\omega - (m+1/2)\Omega + \mathcal{M}_d^F$ for antiparticles. As a result the partition function is (see Sec.~3.3 of I1)
    \be Z =  \left[\prod_{q d} \left(1+e^{-\beta(\omega - (m+1/2)\Omega - \mathcal{M}_d^F)}\right)\right] \left[\prod_{q d} \left(1+e^{-\beta(\omega - (m+1/2)\Omega + \mathcal{M}_d^F)}\right)\right], \label{Zpartf}\ee
    where the first square bracket refers to particles and the second one to antiparticles.
    Then, using  
\be \log Z =   \sum_{q d} \left(\log \left(1+e^{-\beta(\omega - (m+1/2)\Omega - \mathcal{M}_d^F)}\right)
+\log \left(1+e^{-\beta(\omega - (m+1/2)\Omega + \mathcal{M}_d^F)}  \right)\right), \label{logZf}\ee 
one finds that the only non-vanishing averages of products of two annihilation and creation operators are
\bea \langle \gamma^\dagger_{q, d}\gamma_{q', d'}\rangle &=& f_F(\omega - (m+1/2)\Omega - \mathcal{M}_d^F)\delta_{dd'} \delta_{qq'}, \label{dgamma}\\ 
\langle \delta^\dagger_{q, d}\delta_{q', d'}\rangle &=& f_F(\omega - (m+1/2)\Omega + \mathcal{M}_d^F) \delta_{dd'} \delta_{qq'}, \label{ddelta}
\\ 
\langle \gamma_{q, d}\gamma^\dagger_{q', d'}\rangle &=& (1-f_F(\omega - (m+1/2)\Omega - \mathcal{M}_d^F)) \delta_{dd'} \delta_{qq'}, \label{gammad}\\ \langle \delta_{q, d}\delta^\dagger_{q', d'}\rangle &=& (1-f_F(\omega - (m+1/2)\Omega + \mathcal{M}_d^F)) \delta_{dd'} \delta_{qq'}, \label{deltad}\eea
where
\be f_F(x) \equiv \frac1{e^{\beta x}+1} \label{fFdef}\ee
is the Fermi-Dirac distribution. Setting $d=d'$ and $q=q'$ in~(\ref{dgamma}) and~(\ref{ddelta}) one finds the average numbers of fermions and antifermions. More specific formul\ae~without chemical potentials were found by~\cite{Vilenkin1978}.

In the small-temperature limit, which is  typically relevant, for example, in neutron stars, $f_F(x)\simeq \theta(-x)$, where $\theta(x)$ is the Heaviside step function. Therefore, the average numbers go to 1 or 0 in this limit for $\omega - (m+1/2)\Omega  <\pm\mathcal{M}_d^F$ or $>\pm\mathcal{M}_d^F$, respectively, where the plus and minus signs refer to fermions and antifermions, respectively.

Let us recall that in~(\ref{Zpartf}) a single fermion irreducible representation of $\mathcal{G}$ is considered: to obtain the partition function for all  irreducible representations one can take the product of all partition functions of single irreducible representations.

Moreover, using~(\ref{logZf}) and the general expressions~(\ref{avHg}),~(\ref{avJg}) and~(\ref{avQg}), the average values of $H$, $\vec J$  and $Q^a$ turn out to be 
\bea \langle H\rangle &=&  
\sum_{q d} \omega \left(f_F(\omega - (m+1/2)\Omega - \mathcal{M}_d^F)+f_F(\omega - (m+1/2)\Omega + \mathcal{M}_d^F)\right),\label{Hf} \\   \langle  J_z\rangle &=&  
\sum_{q d} \left(m+\frac12\right) \left(f_F(\omega - (m+1/2)\Omega - \mathcal{M}_d^F)+f_F(\omega -(m+1/2)\Omega + \mathcal{M}_d^F)\right), \label{Jzf}\\
\langle Q^a\rangle&=& 
\sum_{q d} \frac{\partial\mathcal{M}_d^F}{\partial\mu_a}\left(  f_F(\omega -(m+1/2)\Omega - \mathcal{M}_d^F)-  f_F(\omega -(m+1/2)\Omega + \mathcal{M}_d^F)\right) \label{avQf}.\eea
In the limit $\Delta p\to0$ (which corresponds to $L\to\infty$) these expressions allow us to compute  the average values of the energy, angular momentum and charges per unit of length in the $z$ direction in terms of integrals rather than sums over $p$. 

We can also go back to the original basis by inverting~(\ref{fromttoD}) and obtain, starting from~(\ref{dgamma})-(\ref{deltad}), that the only non-vanishing averages of pairs of annihilation and creation operators are\footnote{Functions of the matrices $\omega - (m+1/2)\Omega - \mu_a \bar {\cal R}^a$ and $\omega - (m+1/2)\Omega - \mu_a  {\cal R}^a$  can be computed with the spectral decomposition of those matrices as illustrated in Sec.~\ref{Free Scalars}.}
\bea \langle \gamma^\dagger_{qs}\gamma_{q's'}\rangle &=& f_F(\omega - (m+1/2)\Omega - \mu_a \bar {\cal R}^a)_{ss'} \delta_{qq'}, \label{dgammad}\\ 
\langle \delta^\dagger_{qs}\delta_{q's'}\rangle &=& f_F(\omega - (m+1/2)\Omega - \mu_a  {\cal R}^a)_{ss'} \delta_{qq'}, \label{ddeltad}
\\ 
\langle \gamma_{qs}\gamma^\dagger_{q's'}\rangle &=& (1-f_F(\omega - (m+1/2)\Omega - \mu_a \bar {\cal R}^a))_{s's} \delta_{qq'}, \label{gammadd}\\ \langle \delta_{qs}\delta^\dagger_{q's'}\rangle &=& (1-f_F(\omega - (m+1/2)\Omega - \mu_a {\cal R}^a))_{s's} \delta_{qq'}, \label{deltadd}\eea
with $\gamma_{q s} \equiv W_{ds}\gamma_{q, d}$, $\delta_{q s} \equiv W^*_{ds}\delta_{q, d}$.
Corresponding averages for density matrices that also include, unlike~(\ref{rhoRest}), the generators of Lorentz boosts were presented by~\citet{Palermo2021}. 
Here, a closed form is found for density matrices of the form~(\ref{rhoRest}) in the presence of an arbitrary number of chemical potentials. 

Taking into account the expression~(\ref{logZf}) one finds that the bound $\Omega<1/R$ derived in Sec.~\ref{Free Scalars} is also required for the convergence of the fermion averages, so $\Omega <1/R$.

Let us explain now what changes in the case of Majorana fermions in Eq.~(\ref{FreeWexp}). Since in that case one has one type of annihilation operators, $a_{qs}$, rather than two, $c_{qs}$ and $d_{qs}$, only the contribution due to fermions  in~(\ref{rhofref}) is present, while that of antifermions  is absent. This is due to the fact that Eq.~(\ref{FreeWexp}) represents Majorana particles, which coincide with their antiparticles. Correspondingly, $\langle H\rangle$, $\langle J_z\rangle$ and $\langle Q^a\rangle$ will only have the fermion contribution, not the antifermion one.

Some applications and consequences of the results  presented in this Sec.~\ref{fermion Free fields} are discussed in~\citet{Salvio2025c}, where the results were derived.

\section{General interacting theories and path-integral formul\ae}\label{Interacting theories and general path-integral formulas}

So far free particles or quasi-free particles (where in-medium effects are included by effective masses  and effective chemical potentials) have been discussed. Now, we investigate general interacting theories, keeping $\Omega$, $T$  and the $\mu_a$ arbitrary.

 In TFT one is interested in computing thermal Green's functions
 \be  \langle {\cal T}\mathcal{O}_1(x_1) ...  \mathcal{O}_n(x_n) \rangle= \Tr (\rho \mathcal{T} \mathcal{O}_1(x_1) ...  \mathcal{O}_n(x_n)), \label{GreenO} \ee
 where ${\cal T}$ is the time-ordered product and the $\mathcal{O}_i$  are operators involving any kind of fields (scalars, vectors and/or fermions).   These functions are indeed very useful tools to compute the rates of inclusive processes, such as decay or scattering rates
summed over final states or production rates summed over initial states (see Sec.~2 of I1).  Here we investigate the general interacting case for an arbitrary  $n$-point Green's function. 
 
 The $\mathcal{O}_i(x)$ are operators in the Heisenberg picture 
 \be \mathcal{O}_i(t,\vec x) \equiv  \exp(iHt)\,\mathcal{O}_i(0,\vec x)\,\exp(-iHt) \label{OHeis}.\ee
 We take the $\mathcal{O}_i$ in well-defined representations of the Lorentz and internal-symmetry group $\mathcal{G}$. Under the element of $\mathcal{G}$ with parameters $\alpha_a$
 \be \exp(i\alpha_aQ^a)\mathcal{O}_i(x)\exp(-i\alpha_aQ^a) = \exp(i\alpha_aT^a_i)\mathcal{O}_i(x), \label{GonOi} \ee
 where the $T_i^a$ are the generators of the internal-symmetry group in the representation of $\mathcal{O}_i$.
 
For the derivation of the path-integral formul\ae~it will also be convenient to introduce 
 \be \mathcal{\tilde O}_i(t,\vec x) \equiv  \exp(i(H-\vec\Omega\cdot \vec J\,)t)\,\mathcal{O}_i(0,\vec x)\,\exp(-i(H-\vec\Omega\cdot \vec J\,)t) =  \mathcal{D}_i(t\vec\Omega)\mathcal{O}_i(t, R(-t\vec\Omega)\vec x), \label{tildeO}\ee
 where in~(\ref{tildeO}) the repeated index $i$ does not understand a summation.
 Here,  $x^\omega\equiv \{t, R(t\vec\Omega)\vec x\}$ and $\mathcal{D}_i(t\vec\Omega)$  implement the rotation of angle $t\Omega$ around $\vec\Omega$ on the spacetime coordinates and in the Lorentz representation of $O_i$, respectively. The quantity $R(t\vec\Omega)$ is the $3\times3$ orthogonal matrix with unit determinant representing the rotation associated with $t\vec\Omega$. In the right-hand-side of~(\ref{tildeO}) the operator $\mathcal{O}_i$ is computed in $x^{-\omega}\equiv \{t, R(-t\vec\Omega)\vec x\}$. Note that to agree with our definitions of free fields  with angular-momentum wavefunctions $e^{im\phi}$ and  of free angular-momentum operator adopted in Sec.~\ref{Free fields}, the transformation $\vec x\to R(t\vec\Omega)\vec x$ corresponds to $\phi\to\phi-\Omega t$.  Note also  that $\mathcal{\tilde O}_i$ transforms like $\mathcal{O}_i$ under $\mathcal{G}$ because $\mu_a Q^a$ commutes with $H$ and $\vec \Omega \cdot \vec J$. The relation in~(\ref{tildeO}) can be inverted as follows. Consider first
 \be \exp(-i\vec\Omega\cdot \vec J\, t')\,\mathcal{\tilde O}_i(t,\vec{x})\,\exp(i\vec\Omega\cdot \vec J\,t') = \mathcal{D}_i((t+t')\vec\Omega)\mathcal{O}_i(t, R^{-1}((t+t')\vec\Omega)\vec x),\ee
 where $t'$ is an arbitrary time. On the other hand, one also has
 \be \exp(-i\vec\Omega\cdot \vec J\, t')\,\mathcal{\tilde O}_i(t,\vec{x})\,\exp(i\vec\Omega\cdot \vec J\,t')  = \mathcal{D}_i(t'\vec\Omega) \mathcal{\tilde O}_i(t, R^{-1}(t'\vec\Omega)\vec x), \ee
 so, setting $t'=-t$, one finds the inverted relation 
 %XXXXXX\xxx{$1\to i$ in the expression below}
\be \mathcal{O}_i(x_i)= \mathcal{D}_i^{-1}(t_i\vec\Omega) \mathcal{\tilde O}_i(x_i^\omega)\equiv O^\omega_i(x_i^\omega). \label{InTr}\ee

\subsection{Generalized Kubo-Martin-Schwinger conditions}\label{Generalized Kubo-Martin-Schwinger condition}

Let us start by considering $2$-point Green's functions. 
In TFT there are consistency conditions that these functions should satisfy, first found in a finite-temperature theory by~\citet{Kubo1957},~\cite{Martin1959}. These KMS conditions represent an exact (non-perturbative) relation between the two non-time-ordered 2-point functions of a pair of field operators, $\mathcal{O}_1$ and $\mathcal{O}_2$, in a finite-temperature field theory. These conditions were recently extended to include generic values of both the $\mu_a$ and $\Omega$, keeping the  Lorentz and internal-symmetry representations for $\mathcal{O}_1$ and $\mathcal{O}_2$ general (generalized KMS condition) in~\cite{Salvio2026} (see also~\citet{Landsman1987},~\citet{Quiros1994},~\citet{Bellac2011} and~\citet{Ambrus2021} for previous more specific results). So one can apply these conditions to {\it any} TFT with equilibrium density matrix. Let us derive these generalized KMS conditions both in coordinate- and momentum-space. 
 
 \subsubsection{Coordinate space}\label{Coordinate space}
Exploiting the ciclicity of the trace one finds
\bea &&G^<(x_1,x_2)\equiv \langle \mathcal{O}_2(x_2)\mathcal{O}_1(x_1)\rangle \equiv \frac1{Z} \Tr\left(e^{-\beta (H-\vec\Omega  \cdot \vec J - \mu_a Q^a)}\mathcal{O}_2(x_2)\mathcal{O}_1(x_1)\right)\nonumber \\
 &=&  \frac1{Z} \Tr\left(e^{-\beta (H-\vec\Omega  \cdot \vec J - \mu_a Q^a)} e^{\beta (H-\vec\Omega  \cdot \vec J - \mu_a Q^a)}\mathcal{O}_1(x_1)e^{-\beta (H-\vec\Omega  \cdot \vec J - \mu_a Q^a)} \mathcal{O}_2(x_2)\right).\eea 
On the other hand, using~(\ref{GonOi}) and~(\ref{tildeO}) one can write 
\be e^{\beta (H-\vec\Omega  \cdot \vec J - \mu_a Q^a)}\mathcal{O}_1(x_1)e^{-\beta (H-\vec\Omega  \cdot \vec J - \mu_a Q^a)}  = e^{-\beta \mu_a T^a_1} \mathcal{D}_1(-i\beta\vec\Omega)\mathcal{O}_1(t_1-i\beta,R(i\beta \vec \Omega)\vec x_1),\ee 
(note that the $T_1^a$ commute with $\mathcal{D}_1$, as they act on different types of indices), to obtain 
\be \boxed{G^<(x_1,x_2) = e^{-\beta \mu_a T^a_1} \mathcal{D}_1(-i\beta\vec\Omega)G^>(t_1-i\beta,R(i\beta\vec \Omega)\vec x_1,x_2)
,} \label{KMSc}\ee
where $G^>(x_1,x_2)\equiv G^>(t_1,\vec x_1, x_2)\equiv \langle \mathcal{O}_1(x_1)\mathcal{O}_2(x_2)\rangle$.
Eq.~(\ref{KMSc}) represents the desired generalized KMS conditions in coordinate space.

\subsubsection{Analyticity}\label{Analyticity}

The conditions~(\ref{KMSc}) involve  a non-time-ordered 2-point function evaluated at complex values of its arguments, $G^>(t_1-i\beta,R(i\beta\vec \Omega)\vec x_1,x_2)$. One can show that if $G^>(x_1,x_2)$ is a well-defined function  for real $x_1$ and $x_2$ and $\beta\geq0$ then it remains well defined after the substitution $\{t_1,\vec x_1\}\to \{t_1-i\beta,R(i\beta\vec \Omega)\vec x_1\}$.

 To prove this analyticity property let us use the cylindrical coordinates in~(\ref{CyCoo}) to write 
 $$\{x^1_1, x^2_1, x^3_1\}=\{r_1\cos\phi_1, r_1\sin\phi_1, z_1\}, \qquad \{x^1_2, x^2_2, x^3_2\}=\{r_2\cos\phi_2, r_2\sin\phi_2, z_2\}$$
  and neglect irrelevant rotation matrices around the third axis of angles $\phi_1$ and $\phi_2$. One can then express the 2-point function in question in terms of the following spectral decomposition
 \be
 \hspace{-0.4cm}\Tr\left(\rho\,e^{i(Ht-\phi J_3)}\mathcal{O}_1(0,r_1,0,z)e^{-i(Ht-\phi J_3)}\mathcal{O}_2(0,r_2,0,0)\right)=\frac{1}{Z} \sum_{qq'}\langle q'|e^{\beta \mu_a Q^a}\mathcal{O}_1(0,r_1,0,z)|q\rangle\langle q|\mathcal{O}_2(0,r_2,0,0)|q'\rangle e^{i\omega_{q'}(t+i\beta)-im_{q'}(\phi+i\beta\Omega)}e^{-i \omega_q t+im_q \phi}, \hspace{-0.3cm}\label{AnalyStep}\ee
with $t\equiv t_1-t_2$, $\phi\equiv \phi_1-\phi_2$, $z\equiv z_1-z_2$ and the states $|q\rangle$ form a complete set of eigenstates of $H$ and $J_3$, namely $H|q\rangle = \omega_q|q\rangle$ and $J_3|q\rangle = m_q|q\rangle$. As stated above, one is assuming here that this series converges for $t$ and $\phi$ real and $\beta\geq 0$.  Then it keeps converging even after substituting
\be t\to t-i\tau, \qquad \phi\to \phi-i\tau \Omega, \label{subtphi}\ee
with $\tau \in [0,\beta]$. To see this one can use
\be \mathcal{O}_1 = e^{\tau \mu_a T^a_1}e^{-\tau \mu_a Q^a} \mathcal{O}_1e^{\tau \mu_a Q^a}\ee
(which follows from~(\ref{GonOi})) 
at the same time as the substitution in (\ref{subtphi}) and take $|q\rangle$ to be an eigenstate also of $\mu_aQ^a$ with eigenvalue $\mathcal{M}_q$, which can always be done because the $Q^a$ commute with $H$ and $\vec J$. In this way one obtains
\be \frac{e^{\tau \mu_a T^a_1}}{Z} \sum_{qq'}\langle q'|e^{(\beta-\tau) \mu_a Q^a}\mathcal{O}_1(0,r_1,0,z)|q\rangle\langle q|\mathcal{O}_2(0,r_2,0,0)|q'\rangle  \,e^{i\omega_{q'}(t+i(\beta-\tau))-im_{q'}(\phi+i(\beta-\tau)\Omega)}e^{-i \omega_q t+im_q \phi}e^{-\tau(\omega_q-\Omega m_q -\mathcal{M}_q)}\ee 
and, modulo a harmless overall matrix, $\exp(\tau \mu_a T^a_1)$, one reproduces the initial form of the terms of the series with the substitution $\beta\to \beta -\tau$ and the extra exponential factor $\exp(-\tau(\omega_q-\Omega m_q-\mathcal{M}_q))$. For $\tau>0$ this factor generically helps the convergence of the series in order to ensure that the  averages and $Z$  are finite. Since $\tau \in [0,\beta]$, the  analyticity property mentioned above is thus proved.

\subsubsection{Momentum space}\label{Momentum space}

With the result of Sec.~\ref{Analyticity}, we can derive  the generalized KMS conditions for the (partial) Fourier transforms with respect to the coordinates $\xi\equiv \{t, z, \phi\}$:
\be  \tilde G^>(k; r_1, r_2) \equiv \int d^3 \xi \, e^{ik\xi}  \, G^>(\xi; r_1, r_2), \qquad \tilde G^<(k; r_1, r_2) \equiv \int d^3 \xi \, e^{ik\xi}  \, G^<(\xi; r_1, r_2).\ee
Here we are using the fact that $G^>(x_1, x_2)$ and $G^<(x_1, x_2)$ are in fact functions only of $t\equiv t_1-t_2$, $\phi\equiv \phi_1-\phi_2$, $z\equiv z_1-z_2$ and $r_1$ and $r_2$ and so they can be expressed in the form $G^>(\xi; r_1, r_2)$ and $G^<(\xi; r_1, r_2)$, respectively. This is because $\vec\Omega$ breaks the full rotational and space-translational symmetry, but preserves cylindrical symmetry (with axis in the $\vec\Omega$ direction), while time-translational symmetry is preserved. Also, we defined $k\xi \equiv k_0 t - p z -m \phi$.

Using the generalized coordinate-space KMS conditions in~(\ref{KMSc}), one finds
\bea  &&\tilde G^<(k; r_1, r_2) = \int d^3\xi \, e^{ik\xi} \, e^{-\beta \mu_a T^a_1} \mathcal{D}_1(-i\beta\vec\Omega)G^>(t_1-i\beta, R(i\beta\vec \Omega)\vec x_1,x_2) \nonumber \\
&=&e^{-\beta (k_0-m\Omega)} e^{-\beta \mu_a T^a_1} \mathcal{D}_1(-i\beta\vec\Omega) \int dz\int_{-\infty-i\beta}^{+\infty-i\beta} dt \int_{-i\beta \Omega}^{2\pi-i\beta \Omega} d\phi \, e^{ik\xi} \,G^>(\xi; r_1, r_2). \eea
The analyticity discussed in Sec.~\ref{Analyticity} then tells us
 \be \boxed{\tilde G^<(k; r_1, r_2) = e^{-\beta (k_0-m\Omega)} e^{-\beta \mu_a T^a_1} \mathcal{D}_1(-i\beta\vec\Omega) \tilde G^>(k; r_1, r_2),}  \ee
which represents  the generalized KMS conditions in momentum space. This version  of the generalized KMS conditions can be used, among other things, to easily relate absorption and production rates of particles coupled to the rotating plasma~\citep{Altherr1993}.

 \subsection{Scalar path integral}\label{ScalarPI}
 We now derive a useful path-integral representation for $Z$ and general Green's functions, Eq.~(\ref{GreenO}), starting from the simplest case where the fundamental operators are all scalar; the fermion and gauge-field cases will be included in Secs.~\ref{FermionPI} and~\ref{GaugePI}. Let us consider already here, however, a generic number of scalar fields, which we collectively represent by an array\footnote{In order to write more compact expressions, the index of $\Phi$, which $\mathcal{G}$ can act on, is understood. The field $\Phi$ can include here an arbitrary number of irreducible representations of  $\mathcal{G}$.} $\Phi$ of Hermitian field operators.  
 
 Let us recall that $\Phi(x)$ is the field operator in the  Heisenberg picture, $\Phi(t,\vec x)\equiv \exp(iHt)\Phi(0,\vec{x})\exp(-iHt)$. 
 Being Hermitian, $\Phi(0,\vec{x})$ admits a complete set of eigenstates $|\varphi\rangle$, namely $\Phi(0,\vec{x})|\varphi\rangle = \varphi(0,\vec{x}) |\varphi\rangle$, where $\varphi(0,\vec{x})$ is the corresponding eigenvalue.
 
  In order to derive a path-integral expression for~(\ref{GreenO}), instead of using only $\Phi(t,\vec x)$ and the corresponding eigenstates $|\varphi,t\rangle\equiv \exp(iHt)|\varphi\rangle$, it is  convenient to use~$\tilde \Phi(t,\vec x)$ (recall the general definition in~(\ref{tildeO}))
 and its eigenstates
  \be |\tilde\varphi,t\rangle \equiv \exp(i(H-\vec\Omega\cdot \vec J\,)t) |\varphi\rangle, \label{tildevar}\ee
 which contain both $H$ and $\vec\Omega$. Note that $|\tilde\varphi,t\rangle$ is an eigenstate of $\tilde \Phi(t,\vec x)$ with eigenvalue $\varphi(0,\vec x)$.

   We choose a standard normalization of the states in~(\ref{tildevar}) such that 
\be \int \delta\varphi(t) |\tilde\varphi,t\rangle\langle\tilde\varphi,t| = 1, \qquad \mbox{where} \qquad \delta\varphi(t) \equiv \prod_{\vec{x}} d\varphi(t,\vec{x}).\ee
Under $\mathcal{G}$ the field operator $\Phi(x)$ transforms according to the general rule in~(\ref{GonOi}) with the $T_i^a$ set to $\theta^a$ (the generators of $\mathcal{G}$ in the scalar representation). Let us define 
\be \exp(-i\alpha_aQ^a)|\tilde \varphi,t\rangle \equiv |e^{i\alpha_a\theta^a}\tilde\varphi,t\rangle \ee
because this state is an eigenstate of $\tilde\Phi(x)$ with eigenvalue $\exp(i\alpha_a\theta^a) \varphi(0,\vec x)$. So  
\be \langle\tilde \varphi,t| \exp(i\alpha_aQ^a)= \langle e^{i\alpha_a\theta^a}\tilde\varphi,t|. \ee
 Choosing now the basis $\{|\tilde\varphi,t_0\rangle\}$ to represent the trace in~(\ref{GreenO}), where $t_0$ is an arbitrary time, one finds
\bea \langle {\cal T}\Phi(x_1)...\Phi(x_n) \rangle &=& \int \delta \varphi(t_0) \langle\tilde\varphi,t_0|\rho\,  {\cal T}\mathcal{O}_1(x_1) ...  \mathcal{O}_n(x_n)|\tilde\varphi,t_0\rangle \nonumber \\
&=&\frac1{Z}\int \delta \varphi(t_0) \langle\tilde\varphi,t_0|e^{-\beta (H-\vec\Omega  \cdot \vec J - \mu_a Q^a)}  {\cal T}\mathcal{O}_1(x_1) ...  \mathcal{O}_n(x_n)|\tilde\varphi,t_0\rangle \nonumber\\&=&\frac1{Z}\int \delta \varphi(t_0) \langle e^{\beta \mu_a \theta^a} \tilde\varphi,t_0-i\beta|  {\cal T}\mathcal{O}_1(x_1) ...  \mathcal{O}_n(x_n)|\tilde\varphi,t_0\rangle\nonumber\\&=&\frac1{Z}\int \delta \varphi(t_0) \langle e^{\beta \mu_a \theta^a} \tilde\varphi,t_0-i\beta|  {\cal T}\mathcal{\tilde O}_1^\omega(x_1^\omega) ...  \mathcal{\tilde O}_n^\omega(x_n^\omega)|\tilde\varphi,t_0\rangle, \label{ScalarTGF}\eea 
having used~(\ref{InTr}) in the last step.

The thermal Green's functions have been reduced to the corresponding expressions in the absence of rotation, but with Hamiltonian $H-\vec\Omega\cdot \vec J$ and with operators $\mathcal{\tilde O}_i^\omega(x_i^\omega)$ (defined in~(\ref{InTr})) rather than $\mathcal{O}_i(x_i)$. 

Using then the well-known results in the absence of rotation (see e.g.~\citet{Landsman1987} and the references therein), one can now show 
   \be \hspace{-0.7cm}\boxed{\langle {\cal T}\Phi(x_1)...\Phi(x_n) \rangle =\frac1{Z}\int \delta\varphi \, \delta p_\varphi  \, O^\omega_1(x_1^\omega) ... O_n^\omega(x_n^\omega) \exp\left(i \int_C d^4x \left(\dot\varphi(x)p_\varphi(x) - \mathcal{H}^\omega_c(\varphi(x), p_\varphi(x))\right)\right),} \label{PIscalars}\hspace{-0.3cm}\ee 
 where a dot represents a time derivative, $p_\varphi$ is the momentum conjugate to $\varphi$ and the path integration is on all field configurations satisfying  the twisted periodicity condition:
  \be e^{\beta\mu_a \theta^a}\varphi(t_0,\vec{x}) = \varphi(t_0-i\beta,\vec{x}).\label{PerCon}\ee
  Also, $O^\omega_i(x_i^\omega)\equiv \mathcal{D}^{-1}_i(t_i\vec\Omega)O_i(x_i^\omega)$ and $\mathcal{D}_i(t_i\vec\Omega)$  and  the $c$-number fields $O_i(x_i)$ are  obtained by substituting (in the field operators $\mathcal{O}_i(x_i)$) the canonical quantum variables with their eigenvalues after putting all canonical ``coordinates" on the right of all canonical ``momenta".
The integration measures on $\varphi$ and $p_\varphi$ are
  \be \delta\varphi = \prod_x d\varphi(x), \quad \delta p_\varphi= \prod_x d(p_\varphi(x)/(2\pi)).\ee
 The function $\mathcal{H}^\omega_c$ is, in the classical limit, the full classical Hamiltonian density, including the effect of rotation. The defining property of this quantity is
  \be \int d^3x \, \mathcal{H}^\omega_c = H_c- \vec\Omega\cdot \vec J_c, \label{callHc}\ee  
  where $H_c$ and $\vec J_c$ are 
\be H_c(\varphi, p_\varphi)\equiv \frac{\langle\varphi|H|p_\varphi\rangle}{\langle\varphi|p_\varphi\rangle}, \qquad \vec J_c(\varphi, p_\varphi)\equiv \frac{\langle\varphi|\vec J |p_\varphi\rangle}{\langle\varphi|p_\varphi\rangle},  \label{HJc}  \ee 
with $|p_\varphi\rangle$ being the generic eigenstate of the conjugate momentum operator $P_{\varphi}$.     
Moreover,  in the path integral, while the space integral has  no restriction,
 the integral over $t$ is performed on a contour $C$ in the complex $t$ plane that connects  $t_0$ and $t_0-i\beta$ and contains the time components $x_1^0, ... , x_n^0$ of   $x_1, ... , x_n$. This is because we have to include $t_0$, $x_1^0, ... , x_n^0$ and $t_0-i\beta$ in the set of discrete times (that we introduce in deriving the path integral).  
 
 The partition function that appears in the denominator in~(\ref{PIscalars}) is
  \be \boxed{Z =\int \delta\varphi \, \delta p_\varphi  \,  \exp\left(i \int_C d^4x \left(\dot\varphi(x)p_\varphi(x) - \mathcal{H}_c^\omega(\varphi(x), p_\varphi(x))\right)\right),} \label{GenPart}  \ee
  which is nothing but the path integral in~(\ref{PIscalars}) without $O^\omega_1(x_1^\omega) ... O_n^\omega(x_n^\omega)$. Recall that $Z$ can be used to compute the averages of observables, see e.g.~(\ref{avHg}),~(\ref{avJg}) and~(\ref{avQg}).

  The time-ordered product $\mathcal{T}$ should be understood as the product  ordered according to the orientation of $C$: one introduces a parametrization $t(u)$ of $C$, where $u$ is a real parameter such that $t(u)$ proceeds along $C$ with its orientation as $u$ increases. The ordering along the contour is the ordering in $u$. For example, given two points $t$ and $t'$ on $C$, which correspond respectively to $u$ and $u'$, i.e.~$t=t(u)$ and $t'=t(u')$, the Heaviside step function is defined as
\be \theta(t'-t)\equiv \theta(u'-u). \label{contourT}\ee
 The Dirac delta function $\delta$  on the contour $C$ is defined as follows: given two points $t$ and $t'$ on $C$, which correspond respectively to $u$ and $u'$, 
\be \delta(t-t') \equiv \left(\frac{d t}{d u}\right)^{-1} \delta(u-u'). \label{contourDF}\ee
With this definition we have, for a generic function $f$ on $C$,
\be \int_C dt' \delta(t-t') f(t') = f(t) \ee
and 
\be \frac{d}{dt'}\theta(t'-t) = \delta(t'-t). \ee

The contour $C$ can be chosen like in the absence of $\vec\Omega$ and one can recover the imaginary-time or real-time formalism with specific choices. The former is obtained by setting $t_0=0$ and going from $0$ to $-i\beta$ along the imaginary axis~\citep{Matsubara1955}. The latter can be obtained, for example, with the choice of~\citet{Matsumoto1983} (see also Fig.~1 in Sec.~4.1.1 of I1 for a picture in the present notation) and implies a doubling of the degrees of freedom.

  As clear from~(\ref{PIscalars}) and~(\ref{callHc}), in the presence of a non-vanishing average angular momentum one has to substitute  
  $H_c$ with $H_c -\vec \Omega\cdot \vec J_c$ in the path integral. This substitution is nothing but the transformation rule of the classical Hamiltonian from an inertial frame to a frame rotating with angular-velocity vector $\vec \Omega$.      One can, therefore, identify $\vec \Omega$ with the angular-velocity vector of the rotating plasma. One should keep in mind, however, that in TFT $\vec \Omega$ is a thermodynamical quantity, which allows us to compute the average value of the angular momentum (see Eq.~(\ref{avJg})) in {\it any} rotating plasma at thermal equilibrium.

 We can now generate the thermal Green's function by performing functional derivatives of the generating functional 
 \be \boxed{\mathcal{Z} (j) =\frac1{Z} \int \delta\varphi \, \delta p_\varphi  \, \exp\left(i \int_C d^4x \left(\dot\varphi(x)p_\varphi(x) - \mathcal{H}^\omega_c(\varphi(x), p_\varphi(x))+j(x)\varphi(x^\omega)\right)\right).} \label{GenFun}\ee  
 For example, for $\mathcal{O}_1(x_1) ...  \mathcal{O}_n(x_n)=\Phi(x_1)...\Phi(x_n)$, the explicit formula is
 \be \langle {\cal T}\Phi(x_1)...\Phi(x_n) \rangle  = \left.\frac1{ i^n}\frac{\delta^n}{\delta j(x_1) ... \delta j(x_n)}  \mathcal{Z}(j) \right|_{j=0}, \label{ZJ0}\ee
 where here we use the Dirac delta function  on the contour $C$, Eq.~(\ref{contourDF}), to define functional derivatives.

%\paragraph{\bf Momentum integration and Lagrangian path integral for ordinary scalars}
%\label{Momentum integration and Lagrangian path integral for ordinary scalars}

The result in~(\ref{GenFun}) applies to all scalar theories in the presence of generic values of $T$, $\Omega$ and the $\mu_a$. 
 Let us now consider ordinary scalars for which $\mathcal{H}_c$, whose defining property is $H_c =\int d^3 x\mathcal{H}_c$, reads
 \be \mathcal{H}_c = \frac{p_\varphi^2}2 +\frac{(\vec\nabla\varphi)^2}{2}+
 \frac{1}2 
\varphi M^2\varphi+\mathcal{V}(\varphi),  \ee
  where  $\mathcal{V}$ is the potential density minus the mass term $ 
\varphi M^2\varphi/2$. This $\mathcal{H}_c$ can be obtained from the quantum Hamiltonian
  \be H = \int d^3x \left(\frac{P_\varphi^2}2 +\frac{(\vec\nabla\Phi)^2}{2}+ \frac{1}2 
\Phi M^2\Phi+\mathcal{V}(\Phi)\right),\label{Hordinary}\ee
  with $P_\varphi$ being the conjugate-momentum field operator, $P_\varphi|p_\varphi\rangle=p_\varphi|p_\varphi\rangle$. 
  
 To proceed we need the explicit expressions of the  operators $\vec J$ and $\vec P$; one can take
   \be J_k = \epsilon_{ijk}\int d^3x \, (x^j\partial_i\Phi) P_\varphi, ~ \iff ~\vec J= -\int d^3x \, (\vec x \times \vec\nabla\Phi ) P_\varphi , \qquad  P^i = -\int d^3x \, \partial_i\Phi \, P_\varphi.\ee
  Indeed, these expressions lead to the well-known commutators of $H$, $\vec P$ and $\vec J$. Using the second definition in~(\ref{HJc}) one finds
  \be \vec J_c = -\int d^3x \, (\vec x \times \vec\nabla\varphi ) p_\varphi. \ee

In this class of theories one can perform the integration over $p_\varphi$, as we now show.
  In the present case $\mathcal{H}_c^\omega$ only features terms quadratic, linear  and constant in $p_\varphi$ so that the  integration over $p_\varphi$ can be easily done with the stationary-point method. Including the effect of $\vec\Omega$, the argument of the exponential in~(\ref{GenFun}) is stationary at
  \be p_\varphi = \dot\varphi-\vec\Omega\cdot  \vec x \times \vec\nabla\varphi.  \ee
  Using this expression in integrating over $p_\varphi$ in~(\ref{GenFun}), one finds the following Lagrangian path integral:
   \be\boxed{\mathcal{Z}(j) = \frac1{``j\to 0"} \int \delta\varphi \exp\left(iS_C(\varphi)+i\int_C d^4x j(x)\varphi(x^\omega)\right),}  \label{GenFunLag} \ee
   where the denominator ``$j \to0$" is the numerator for $j \to0$,
   \be \boxed{S_C(\varphi) = \int_C d^4x \left(-\frac12 \varphi (\Box_\omega+M^2)\varphi - \mathcal{V}(\varphi) \right)} \label{SC}\ee
 is the action computed with the contour $C$ and the operator $\Box_\omega$
  is the d'Alembertian operator after the substitution $\partial_t \to \partial_t -\vec\Omega\cdot \vec x\times\vec\nabla$.  
 
 It is useful to note that $\vec\Omega$ only appears in the  quadratic action. 
 This implies that, in perturbation theory, only the thermal propagators are modified by $\vec\Omega$, the vertices are unmodified. Indeed, for an interaction term without time derivatives, say $\int_C d^4x  \mathcal{V}(\varphi(x))$, one can perform a change of variables (a rotation of angle $t\Omega$ around $\vec\Omega$, which has unit Jacobian\footnote{The Jacobian is $\cos^2(t\Omega)+\sin^2(t\Omega)$, which is equal to 1 for any real $t$ and its analytic continuation to complex $t$ remains 1.})
  in the spatial integration  and rewrite it as $\int_C d^4x  \mathcal{V}(\varphi(x^\omega))$ so that, in the perturbative expansion, one ends up computing   
 correlators with field insertions of the form $O^\omega(x_1^\omega)...O^\omega(x_n^\omega)$. Also, performing the perturbative expansion of $\mathcal{Z}(j)$ and using~(\ref{ZJ0}) one  finds that the vertices depend neither on $\Omega$ nor on the $\mu_a$.  For the computation of the vertices one can use well-known results from the literature (see e.g.~\citet{Bellac2011} and~\cite{Landsman1987}); on the other hand, the thermal propagators will be computed below in  Sec.~\ref{ScalarProp}, including the effect of $\vec\Omega$ and generic $\mu_a$. This will provide a practical recipe to perform perturbation theory in the presence of both $\vec\Omega$ and  $\mu_a$. Some applications with $\Omega\neq0$ can be found in~\cite{Salvio2025b},~\cite{Siri2025} and~\cite{Kuboniwa2026}.

 Eq.~(\ref{SC}) shows that, as geometrically clear, $\vec\Omega$ breaks rotational and space-translational symmetry down to cylindrical symmetry (with axis in the $\vec\Omega$ direction), while time-translational symmetry is preserved. In cylindrical coordinates~(\ref{CyCoo})
  \be \vec\Omega\cdot \vec x\times\vec\nabla = \Omega \partial_\phi \ee
 and the operator in the quadratic action reads
 \be \Box_\omega+M^2
 =(\partial_t-\Omega\partial_\phi)^2 - \partial_r^2 -\frac1r \partial_r-\frac1{r^2}
 \partial_\phi^2 - \partial_z^2 +M^2. \label{Opomega} \ee
 So  the term due to $\vec \Omega$ does not contribute positively to the Euclidean action, which one has to consider in the imaginary-time formalism. 
 
 This situation is reminiscent of the well-known sign problem due to the $\mu_a$.
 To understand more clearly this point one can consider the following change of variables in the path integral:
 \be d(t,\vec x)\equiv e^{-i\mu_a\theta^a t}\varphi(t,\vec x), \ee
 which can be viewed as the action of a time-dependent element of $\mathcal{G}$ and satisfies, because of~(\ref{PerCon}), $$d(t-i\beta,\vec x) = d(t,\vec x).$$ The generating functional in~(\ref{GenFunLag}) is then left invariant except the fact that one has to impose this simpler periodicity condition and perform the   following substitutions:
 \be \partial_t \to \partial_t+i\mu_a\theta^a, \qquad j(t,\vec x) \to e^{i\mu_a\theta^a t} j(t,\vec x). \ee
So   $S_C$ in~(\ref{SC})
reads  
\be S_C=\int_C d^4x \left\{ -\frac12d\left[(\partial_t -\Omega\partial_\phi+i\mu_a\theta^a)^2-\vec\nabla^2+M^2\right] d -\mathcal{V}(d) \right\}. \nonumber
\ee
As a result, both $\mu_a$ and $\Omega$ give a contribution to the Euclidean action that is not bounded from below (recall that the $\theta^a$ are purely imaginary and antisymmetric and so Hermitian). 

To avoid this problem entirely one can substitute $\mu_a \to i \mu^E_a$ and  $\Omega\to i\Omega_E$, treat $\mu^E_a$ and $\Omega_E$ as real parameters to evaluate the path integral in Euclidean spacetime and finally perform the analytic continuation $\mu^E_a\to -i\mu_a$, $\Omega_E\to -i \Omega$ to obtain physically relevant results.

It is interesting to note, though, that these substitutions are not  necessary in some cases. Consider, for example, $\mu_a=0$ and rewrite the operator in~(\ref{Opomega}) as
\be \Box_\omega+M^2 =\partial_t^2-2\Omega\partial_\phi\partial_t +(- \partial_r^2 -\frac1r \partial_r-\frac1{r^2}
 \partial_\phi^2 +\Omega^2\partial_\phi^2) - \partial_z^2 +M^2. \ee 
 The first term is the familiar $-\partial_\tau^2$ at imaginary time $\tau$ and contributes positively to the Euclidean action. The same is true for $- \partial_z^2$ and $M^2$ if its eigenvalues are non negative.  The second term becomes $-2i\Omega\partial_\phi\partial_\tau$ and gives a harmless imaginary contribution to the  Euclidean action. Finally, note that the operator in the brackets has eigenvalues $\alpha^2-m^2\Omega^2$. While the contribution of $\Omega$ is negative as already stated, in Sec.~\ref{Free Scalars} we have seen, putting the system in a finite cylinder of radius $R$ and height $L$, that $1/R>\Omega$ and also $\alpha\geq |m|/R$~\citet{Vilenkin1980} so that $\alpha^2-m^2\Omega^2>0$ (a positive contribution to the Euclidean action). So one can see that the negative contribution of $\Omega$ is compensated by other positive contributions. 
 
 The conclusion is, therefore, that the real part of the Euclidean action is bounded from below, at least for $\mu_a=0$.
 This opens up the possibility to perform non-perturbative calculations on the lattice in the imaginary-time formalism, even in the presence of a non-zero average angular momentum. Another way to see why this result holds is by noting that going from a non-rotating to a rotating frame corresponds to substituting $H\to H-\vec\Omega \cdot \vec J$, but the convergence of the partition function requires $H-\vec\Omega \cdot \vec J$ and the real part of the Euclidean action to be bounded from below. 
 Moreover, at least on a finite lattice (for which the renormalization procedure is not yet necessary),  the convergence of the ensemble averages of all operators implies the convergence of all  Euclidean thermal Green's functions.
 
 \subsection{Scalar thermal propagators} \label{ScalarProp}
 
The thermal propagator for scalar operators $\Phi_s$ is defined by 
\bea \langle {\cal T}\Phi_s(x_1)\Phi_{s'}(x_2)\rangle&\equiv&\theta(t_1-t_2)\langle \Phi_s(x_1)\Phi_{s'}(x_2)\rangle+\theta(t_2-t_1)\langle \Phi_{s'}(x_2)\Phi_s(x_1)\rangle, \label{ThermalProp}
 \eea
 where $t_{1,2}$ are the time components of the spacetime points $x_{1,2}$.  The thermal (free) propagator is an essential ingredient to perform perturbation theory. 
 
 The easiest way to determine it is by using the averages of the product of pairs of annihilation and creation operators determined in Sec.~\ref{Free Scalars}. In this way one obtains, for the  non-time-ordered 2-point function of a field $\Phi$ in an arbitrary irreducible representation of $\mathcal{G}$,
\bea G_{ss'}^>(x_1,x_2)\equiv \langle \Phi_s(x_1)\Phi_{s'}(x_2)\rangle   &=& \SumInt_q~\,\left[ (1+f_B(\omega - m\Omega - \mu_a R^a))_{s's}\varphi_q(\vec x_1) \varphi_q^*(\vec x_2)e^{-i\omega (t_1-t_2)} \right.\nonumber \\ 
&& \hspace{2cm}\left.  + f_B(\omega - m\Omega - \mu_a R^a)_{ss'} \varphi_q^*(\vec x_1) \varphi_q(\vec x_2)e^{i\omega (t_1-t_2)} \right],  \label{Wf} \eea
where $\varphi_q$ is a shorthand notation for $\varphi_{\omega p m}$.
We recall that $R^a$ are the generators of $\mathcal{G}$ in the generic irreducible representation under study. 
 Once $\mathcal{G}$ and the irreducible representation are specified one can determine  $f_B(\omega - m\Omega - \mu_a R^a)$ appearing in~(\ref{Wf}) by determining  the eigenvalues and the projectors of $\mu_af^a$ as explained around Eq.~(\ref{SpDec}).  
The other non-time-ordered 2-point function appearing in the thermal propagator in~(\ref{ThermalProp}), i.e.
\be G_{ss'}^<(x_1,x_2)\equiv \langle \Phi_{s'}(x_2)\Phi_{s}(x_1)\rangle \label{Gback}\ee 
can  be easily extracted from $G_{ss'}^>(x_1,x_2)$ using the property $G_{ss'}^<(x_1,x_2) = G_{s's}^>(x_2,x_1)$. One can then easily combine $G^>$ and $G^<$ to obtain the thermal propagator using its definition in~(\ref{ThermalProp}).
One can express~(\ref{ThermalProp}) in terms of an integral over four momentum variables $\{k_0, \alpha, p, m\}$ by integrating over $\alpha$ and $p$  instead of $\omega$ and $p$ using $d\omega=\alpha d\alpha/\omega$ in the domain $\alpha\in[0,\infty)$ and $p\in(-\infty, +\infty)$ and  the formula
\be \theta(t) = \lim_{\varepsilon\to 0^+}\frac1{2\pi i} \int_{-\infty}^{+\infty} \frac{dk_0}{k_0 -i\varepsilon} e^{ik_0 t}.  \label{4Mom}\ee%?However, in performing perturbative calculations in TFT it is typically more convenient to use $G^>$ and $G^<$ rather than the full thermal propagator (I1).

\citet{Vilenkin1980} discussed a different 2-point function: the thermal average of the time-ordered product (at different spacetime points) of the Matsubara field operator for vanishing $\mu_a$. This does not coincide with $\Phi_s$ but with the field $\tilde \Phi_s$ (recall the general definition in~(\ref{tildeO})), after the substitution  $it\to \tau$, where $\tau$ is the imaginary time. 

The expression of the thermal propagator given in this section, together with the information that the vertices do not depend on $\Omega$ and the $\mu_a$, has been applied to particle production for $\Omega\neq$~\citep{Salvio2025b}.

The thermal propagator in~(\ref{ThermalProp})-(\ref{Gback}) can be equivalently computed through functional methods and the generalized KMS conditions, used as boundary conditions, as we now show. We do so because this method will be easily adapted in Sec.~\ref{propGauge} to compute the gauge-field and Faddeev-Popov ghost thermal propagators in some gauges. 

Since we  consider scalars in an arbitrary irreducible representation of the internal-symmetry group, the generating functional is given by~(\ref{GenFunLag}) and~(\ref{SC}) with the substitution $M^2\to\mu^2$, where $\mu$ is the mass of the scalars in the irreducible representation in question.

Performing a rotation of angle $t\Omega$ around $\vec\Omega$ of the spatial integration variable (which has unit Jacobian)
  to rewrite the coupling with the external current in the action as 
\be \int_C d^4x j(x)\varphi(x^\omega) = \int_C d^4x j(x^{-\omega})\varphi(x),  \label{ChangSpI}\ee
 one finds that the stationary point $\varphi_0$ of the argument of the exponential in the path integral satisfies 
\be (\Box_\omega+\mu^2)\varphi_0 = j(x^{-\omega}). \label{stationaryScalar}\ee
The solution of this equation is
\be \varphi_0(x) = -\int_C d^4y D(x,y)  j(y^{-\omega}),\ee
where $D$ here is the Green's function of  $-(\Box_\omega+\mu^2)$:
\be -(\Box_\omega^x+\mu^2) D(x,y)  = \delta(x-y) \label{EqGFD} \ee
(the differential operator $\Box_\omega^x$ is by definition $\Box_\omega$ acting on the spacetime variable $x$). Since the ``matrix" with elements $(\Box_\omega^x+\mu^2)\delta(x-y)$ is symmetric (exchanging $x\leftrightarrow y$ does not change it) one finds the symmetry property $D_{ss'}(x,y)=D_{s's}(y,x)$. 

Computing the path integral with the stationary-point method, one can then easily show  that the thermal propagator is 
\be \langle {\cal T}\Phi_s(x_1)\Phi_{s'}(x_2)\rangle=iD_{ss'}(x_1^\omega,x_2^\omega). \ee
The fact that $D$ here is computed in $\{x_1^\omega,x_2^\omega\}$ rather than in $\{x_1,x_2\}$ is because the external current  in~(\ref{ChangSpI}) and~(\ref{stationaryScalar}) is computed in $x^{-\omega}$ rather than in $x$. This leads to a considerable simplification:
\be -(\Box^x+\mu^2) D(x^\omega,y^\omega)  = \delta(x-y),\label{EqGFDo} \ee
with $\Box^x$ being the d'Alembertian operator acting on the spacetime variable $x$.
So the equation that the thermal propagator satisfies does not depend on the $\mu_a$ and $\Omega$. This dependence will be introduced by the generalized KMS conditions.  

The fact that $iD(x_1^\omega,x_2^\omega)$ is the thermal  propagator implies that $D(x_1,x_2)$ can in fact, just like $G^>(x_1, x_2)$ and $G^<(x_1, x_2)$, be expressed in cylindrical coordinates in terms only of $t\equiv t_1-t_2$, $\phi\equiv \phi_1-\phi_2$, $z\equiv z_1-z_2$ and $r_1$ and $r_2$. Then 
\be \langle {\cal T}\Phi_s(x_1)\Phi_{s'}(x_2)\rangle=iD_{ss'}(\xi^\omega; r_1, r_2), \ee
where $\xi^\omega\equiv \{t, z, \phi-\Omega t\}$ and Eq.~(\ref{EqGFDo}) becomes
\be \left(-\partial_t^2+\partial_{r_1}^2+\frac1{r_1}\partial_{r_1}+\frac1{r_1^2}\partial_{\phi}^2+\partial_z^2-\mu^2\right)D(\xi^\omega; r_1, r_2) = \delta(t)\delta(\phi)\delta(z)\frac{\delta(r_1-r_2)}{r_1}. \ee
We can then write 
\be D(\xi^\omega; r_1, r_2) = \sum_{m=-\infty}^{+\infty}\frac{e^{im\phi}}{2\pi}\int_{-\infty}^{\infty}\frac{dp}{2\pi}e^{ipz}\int_0^\infty d\alpha\alpha  J_m(\alpha r_1)J_m(\alpha r_2) D_s(t,p,m,\alpha), \label{Dxio}\ee
where $D_s$ is some function of $t, p, m$ and $\alpha$.
One can do so because one can always express the inverse of a diagonalizable operator as the sum over all the  eigenvalues of the eigenvectors times their ``bras" (in Dirac notation) times the inverse of the eigenvalues. Imaginary values of $\alpha$ are excluded because $D(\xi^\omega; r_1, r_2)$ should be a solution of the homogeneous version of~(\ref{EqGFDo}) when $t\neq0$.
Using the completeness relation of the Bessel functions
\be \int_0^\infty d\alpha\alpha  J_m(\alpha r_1)J_m(\alpha r_2) = \frac{\delta(r_1-r_2)}{r_1}, \ee
and their closure equation one obtains 
\be \left(-\partial_t^2-\omega^2(\alpha,p)\right)D_s(t,p,m,\alpha) = \delta(t), \label{EqDs} \ee 
with $\omega(\alpha,p)\equiv \sqrt{\mu^2+\alpha^2+p^2}$.
Using now the definition of the time-ordered product ${\cal T}$ one can write
 \be D_s(t,p,m,\alpha) = D^>_s(t,p,m,\alpha)\theta(t)+D^<_s(t,p,m,\alpha)\theta(-t)\ee 
 to obtain (inserting in~(\ref{EqDs}))
 \be D_s^{>(<)}(t,p,m,\alpha) = D_p^{>(<)}(p,m,\alpha)e^{-i\omega(\alpha,p) t}+D_n^{>(<)}(p,m,\alpha)e^{i\omega(\alpha,p) t}, \ee
 \be D_p^>=D_p^<-\frac{i}{2\omega}, \qquad D_n^>=D_n^<+\frac{i}{2\omega},\label{PropStep2}\ee 
 where $D_p^{>(<)}$ and $D_n^{>(<)}$ are functions of $p, m$ and $\alpha$.
 
 The coefficients $D_p^{>(<)}$ and $D_n^{>(<)}$ can now be determined using the generalized KMS conditions in Eq.~(\ref{KMSc}), which imply, recalling that we are dealing with scalars,
 \be D_p^< = e^{-\beta (\omega-m\Omega+\mu_aR^a)}D_p^>, \qquad D_n^< = e^{\beta (\omega+m\Omega-\mu_aR^a)}D_n^>. \label{PropStep3} \ee 
  Combing~(\ref{PropStep2}) and~(\ref{PropStep3}) then gives 
 \be D_p^> = -\frac{i}{2\omega} (1+f_B (\omega-m\Omega+\mu_aR^a)), \qquad D_n^> = -\frac{i}{2\omega} f_B (\omega+m\Omega-\mu_aR^a),\ee 
 which leads precisely to~(\ref{Wf}).

\subsection{Fermion path integral}\label{FermionPI}

Here we obtain the path-integral formula for the thermal Green's functions in~(\ref{GreenO})
   where the $\mathcal{O}_i$ can involve the fermion fields $\Psi_s$. Such a formula, in the presence of $\Omega$ and $\mu_a$ can be derived by combining the corresponding discussion in the absence of $\Omega$ and $\mu_a$ of I1 with the derivation of the path-integral formula of thermal Green's functions for purely scalar theories of Sec.~\ref{ScalarPI}, as it is now illustrated.
   
   First, one groups together all fermion annihilation operators ($c_{qs}$ and $d_{qs}$ for Dirac fermions and $a_{qs}$ for Majorana fermions in~(\ref{FreeWexp})) in a single $a_i$. Here $i$ is a collective index running over both particles and antiparticles (if they are distinct from the particles) as well as all values of $q$ and $s$. One can then introduce right, $|\eta\rangle$, and left, $\langle\eta|$, ``eigenstates" of $a_i$ and $a_i^\dagger$, respectively: 
   \be a_i |\eta\rangle = \eta_i |\eta\rangle, \qquad \langle\eta| a_i^\dagger =  \langle\eta|\eta_i^*, \ee
   where $\eta_i$ and $\eta_i^*$ are Grassmann variables satisfying
   \be \{\eta_i,\eta_j\} =\{\eta_i^*,\eta_j^*\} =\{\eta_i,\eta_j^*\}  = 0, \quad \{\eta_i,a_j\}= \{\eta_i,a^\dagger_j\}=  \{\eta^*_i,a_j\} = \{\eta^*_i,a^\dagger_j\}= 0.  \ee

Then the combination of Sec.~4.2.1 of I1 with the derivation of the path-integral formula of thermal Green's functions for purely scalar theories of Sec.~\ref{ScalarPI} leads to 
\be
\hspace{-0.9cm} 
\boxed{\langle \mathcal{T}\mathcal{O}_1(x_1) ... \mathcal{O}_n(x_n)\rangle 
= \frac1{Z} \int \delta\eta^*\delta\eta \exp\left(\int_C dt\left(-\eta^*(t)\dot \eta(t)-iH^\omega_c(\eta^*(t),\eta(t)\right) \right)O_1^{\omega}(x_1^\omega) ... O_n^{\omega}(x_n^\omega),} \hspace{-0.5cm} \label{PIGFF} 
\ee
where 
\be H^\omega_c(\eta^*,\eta') \equiv \frac{\langle\eta|H^\omega(a^\dagger,a) |\eta' \rangle}{\langle\eta|\eta' \rangle}, \quad  \mbox{with} \quad H^\omega \equiv H-\vec\Omega\cdot \vec J \ee
and the $O_i(x_i)$ are the $c$-number fields obtained by substituting (in the field operators $\mathcal{O}_i(x_i)$) the annihilation and creation operators with  $\eta$ and $\eta^*$, respectively, after putting all annihilation operators on the right of all creation operators. 
Moreover, 
\be\delta\eta^*\delta\eta \equiv \prod_t d\eta^*(t) d\eta(t). \ee 
This formula specifies the integration measure to compute the partition function $Z$ (see e.g.~\cite{Zhu2025} for applications with $\Omega\neq0$).
The partition function $Z$ in~(\ref{PIGFF}) is just the numerator in~(\ref{PIGFF}) for $O_1(x_1) ... O_n(x_n)\to 1$:
\be
\hspace{-0.9cm} 
\boxed{Z
=  \int \delta\eta^*\delta\eta \exp\left(\int_C dt\left(-\eta^*(t)\dot \eta(t)-iH^\omega_c(\eta^*(t),\eta(t)\right) \right).} \hspace{-0.5cm} \label{PIZF} 
\ee
Recall that $Z$ can be used to compute the averages of observables, see e.g.~(\ref{avHg}),~(\ref{avJg}) and~(\ref{avQg}).
The integration in~(\ref{PIGFF}) and~(\ref{PIZF}) is subject to the twisted antiperiodic boundary conditions, 
\be \eta(t_0-i\beta)=-\exp(\beta\mu_aT^a)\eta(t_0), \quad  \eta^*(t_0-i\beta)=-\exp(\beta\mu_a\bar T^a)\eta^*(t_0),\label{ABC}\ee
where the matrices $T^a$ act on the species index of the $a_i$ (and thus of the $\eta_i$) as follows:
\be \exp(i\alpha_aQ^a)a\exp(-i\alpha_aQ^a) = \exp(i\alpha_a T^a)a  \ee 
and $\bar T^a \equiv -(T^a)^*$. In other words, the $T^a$ are the generators of $\mathcal{G}$ in the representation of the $a_i$ (and thus of the $\eta_i$). In the case of Majorana fermions, for which $t^a=\bar t^a\equiv-(t^a)^*$, one takes $T^a=t^a$, while for Dirac fermions  (recalling that the $a_i$ represent all annihilation operators, $c_{qs}$ and $d_{qs}$) 
one can write the $T^a$ in a box-diagonal form, $T^a = $ diag$(t^a,\bar t^a)$, where the first block corresponds to the particles and the second one to the antiparticles. 
The formula in~(\ref{PIGFF}) represents the path-integral formula for general operators $\mathcal{O}_1, ... ,\mathcal{O}_n$ involving fermion fields, whose dynamics is dictated by an arbitrary Hamiltonian. 

 Like in the purely scalar theories, to account for a non-vanishing average angular momentum one has to substitute  
  $H_c$ with $H_c -\vec \Omega\cdot \vec J_c$ in the path integral, which is nothing but the transformation rule of the classical Hamiltonian from an inertial frame to a frame rotating with angular-velocity vector $\vec \Omega$.      Once again we see we can identify $\vec \Omega$ with the angular-velocity vector of the rotating plasma.

Adapting again the discussion of I1 (done with $\Omega=0$ and $\mu_a=0$) to the presence of general $\Omega$ and $\mu_a$, the path integral in~(\ref{PIGFF}) can then be written in terms of $c$-number Grassmann fermion fields  $\psi$ and $\bar \psi$ as follows\footnote{For Dirac fermions one can easily show~(\ref{PIGFFFo}) through the orthogonality conditions in~(\ref{OrtoF}). For the Majorana fermions in~(\ref{FreeWexp}) the proof is complicated by the fact that in general the orthogonality conditions in~(\ref{OrtoFX}) and~(\ref{OrtoFY}) do not include the vanishing of $\int d^3 x \, \bar Y_{q'}(x)X_{q}(x)$. Indeed, this integral is non zero for some $q$ and $q'$  because the $X_q$ and $Y_{q}$ can be viewed as two different basis in the same linear space of functions. In particular it can be non zero only for $\sigma'=\sigma$, $p'=-p$, $m'+1/2=-m-1/2$ and $|\vec p'|=|\vec p|$ (note that $|\vec p'|=|\vec p|$ follows from the fact that both $X_q$ and $Y_q$ have the same absolute value of the energy, $\omega'=\omega$, and  are solutions of the same Klein-Gordon equation%can be shown by using the explicit expression of $\phi_j$ in~\cite{Ambrus:2015lfr} and the closure relation of cylindrical Bessel functions
). The difference with the Dirac fermions is that you need both $X_q$ and $Y_{q}$ to form a solution of the field equations and you cannot express $H=i\partial_t$ in terms of a linear and homogeneous operator acting on the space of two-dimensional spinors. However, the non-vanishing of that integral produce the following terms in $\int_C d^4 x \,   \psi^\dagger(x)\dot\psi(x)$:
\be \SumInt_{q'}\SumInt_q\int_C d^4x\left[\bar X_{q'}(0,\vec x) Y_q(0,\vec x) \alpha^*_{q'}\dot\alpha^*_q+\bar Y_{q'}(0,\vec x) X_q(0,\vec x) \alpha_{q'}\dot\alpha_q\right], \label{ExtraPI}\ee 
where the $\alpha_q$ and $\alpha_q^*$ are the Grassmann variables corresponding to $a_{qs}$ and $a^\dagger_{qs}$, respectively, with species index $s$ understood. Now,~(\ref{YfromX}) (together with the conditions $\sigma'=\sigma$ and $|\vec p'|=|\vec p|$) implies that $\bar X_{q'}(0,\vec x) Y_q(0,\vec x)$ and $\bar Y_{q'}(0,\vec x) X_q(0,\vec x)$ are antisymmetric in $q\leftrightarrow q'$, while an integration by parts over time shows that $\int_Cdt\alpha^*_{q'}\dot\alpha^*_q$ and $\int_Cdt\alpha_{q'}\dot\alpha_q$ are symmetric because $\alpha_q$ and $\alpha_q^*$ are  Grassmann variables. So~(\ref{ExtraPI}) vanishes.}
\be  \boxed{\langle \mathcal{T}\mathcal{O}_1(x_1) ... \mathcal{O}_n(x_n)\rangle = \frac1{``O_i\to 1"} \int \delta\bar\psi\delta\psi\exp\left(\int_C d^4x\left[-\psi^\dagger(x)\dot \psi(x)-i\mathcal{H}^\omega_c(\bar\psi(x),\psi(x))\right] \right)O_1^\omega(x_1^\omega) ... O_n^\omega(x_n^\omega), }\label{PIGFFFo} \ee
subject to the antiperiodic boundary conditions,
\be \psi(t_0-i\beta,\vec x)=-\exp(\beta\mu_at^a)\psi(t_0,\vec x), \quad \bar \psi(t_0-i\beta,\vec x)=-\bar\psi(t_0,\vec x)\exp(-\beta\mu_at^a), \label{AntiPpsi}\ee
where $\bar \psi\equiv \psi^\dagger \gamma^0$ for Dirac spinors and $\bar \psi\equiv \psi^\dagger$ for Weyl spinors.
In~(\ref{PIGFFFo}),  $\mathcal{H}^\omega_c$ is the full classical Hamiltonian density, including the effect of rotation. The defining property of this quantity is
  \be \int d^3x \, \mathcal{H}^\omega_c = H_c^\omega =-\vec\Omega\cdot\vec J_c+ \int d^3x \, \mathcal{H}_c, \label{callHc2}\ee
  where $\vec J_c$ is defined through 
  \be \vec J_c(\eta^*,\eta') \equiv \frac{\langle\eta| \vec J(a^\dagger,a) |\eta' \rangle}{\langle\eta|\eta' \rangle} \ee
  and 
  $\mathcal{H}_c$ is the corresponding classical Hamiltonian density in the absence of rotation. For Dirac spinors 
  %Using the well-known formula for the angular momentum operator for fermions, 
  \be \vec J_c = \int d^3x \psi^\dagger(x)\left[\vec x\times(-i\vec\nabla)+\frac{\vec\sigma_4}{2}\right]\psi(x),\label{DiracJc}\ee 
  where the three components of $\vec\sigma_4$ are the following $4\times 4$ matrices,
  \be \vec\sigma_4 = \left(\bac \sigma^{23} \\ \sigma^{31}\\ \sigma^{12} \ea \right), \quad \mbox{with} \quad \sigma^{\mu\nu} \equiv\frac{i}2 [\gamma^\mu,\gamma^\nu]. \ee 
  For Weyl spinors
   \be \vec J_c = \int d^3x \psi^\dagger(x)\left[\vec x\times(-i\vec\nabla)+\frac{\vec\sigma}{2}\right]\psi(x). \label{WeylJc}\ee
   The formul\ae~in~(\ref{DiracJc}) and~(\ref{WeylJc}) are valid in any theory where the dependence of the Lagrangian density on the time derivative of the spinor field is identical to that in~(\ref{freeLagf}) and~(\ref{freeLagW}). 
Then, in such a theory, one can write 
\be \langle \mathcal{T}\mathcal{O}_1(x_1) ... \mathcal{O}_n(x_n)\rangle = \frac1{``O_i\to 1"} \int \delta\bar\psi\delta\psi\exp\left(i\int_C d^4x\, \mathscr{L}_\omega(\bar\psi(x),\psi(x))\right)O_1^\omega(x_1^\omega) ... O_n^\omega(x_n^\omega), \label{PIGFFF} \ee
where the full Lagrangian density including the effect of rotation, $\mathscr{L}_\omega$, is 
\be \mathscr{L}_\omega = \mathscr{L} + \psi^\dagger(x)\,\,\vec\Omega\cdot\left[\vec x\times(-i\vec\nabla)+\frac{\vec\sigma_c}{2}\right]\psi(x), \ee 
with $\mathscr{L}$ being the full Lagrangian density in the absence of rotation and $\vec\sigma_c=\vec\sigma_{4}$ for Dirac spinors and $\vec\sigma_c=\vec\sigma$ for Weyl spinors.

Like for purely scalar theories, $\vec\Omega$ only appears in the  quadratic action, which implies that, in perturbation theory, only the propagators are modified by $\vec\Omega$. The vertices are unmodified by $\vec\Omega$ and (using the same argument for scalars) the $\mu_a$. For the computation of the vertices one can then use, also for fermions, well-known results from the literature (see e.g.~\citet{Bellac2011} and~\cite{Landsman1987}); on the other hand, the thermal propagators will be computed below, including the effect of $\vec\Omega$ and  $\mu_a$. 

As usual these Green's functions can be obtained by taking functional derivatives of
\be \mathcal{Z}(\bar\kappa,\kappa) = \frac1{``\{\bar\kappa,\kappa\}\to0"} \int \delta\bar\psi\delta\psi
 \exp\left(i\int_C d^4x\, \mathscr{L}_\omega(\bar\psi(x),\psi(x))+i\int_Cd^4x(\bar\kappa(x)\psi^\omega(x^\omega)+\bar\psi^\omega(x^\omega)\kappa(x))\right) \nonumber\ee 
 with respect to the Grassmann sources $\kappa$ and $\bar\kappa$.

As already mentioned, to obtain the path integral with $\vec \Omega\neq0$ one can start from the path integral with $\vec \Omega=0$ and substitute there the classical Hamiltonian in the non-rotating frame with that in the rotating frame with angular-velocity vector $\vec \Omega$.   As a result, if scalar fields are also introduced, the effective scalar Euclidean action obtained by integrating out these fermions must have a real part that is bounded from below when it is so for $\vec\Omega=0$ (see the argument at the end of Sec.~\ref{ScalarPI}).
This extends the proof given at the end of Sec.~\ref{ScalarPI} to all scalar theories, including those obtained by integrating out fermion fields.

\subsection{Fermion thermal propagators}\label{Fermion thermal propagator}

For Dirac fermions the thermal propagator is defined by
\be  \langle {\cal T} \Psi_s(x_1)\bar \Psi_{s'}(x_2)\rangle = \theta(t_1-t_2)\langle \Psi_s(x_1)\bar \Psi_{s'}(x_2)\rangle  -\theta(t_2-t_1)  \langle \bar \Psi_{s'}(x_2)\Psi_s(x_1)\rangle, \label{Tpropf}\ee
where the spinor indices are understood and  $\bar \Psi \equiv \Psi^\dagger \gamma^0$.
Using now the fermion field expansion in~(\ref{FreeFexp}) and the average of pairs of annihilation and creation operators in~(\ref{dgammad})-(\ref{deltadd}) one obtains for the non-time-ordered 2-point functions:
\bea   \hspace{-1cm}S_{ss'}^>(x_1,x_2)  \equiv \langle \Psi_s(x_1)\bar \Psi_{s'}(x_2)\rangle=\SumInt_q \left\{\mathcal{U}_q(x_1)\mathcal{\bar U}_q(x_2)(1-f_F(\omega - (m+1/2)\Omega - \mu_a \bar {\cal R}^a))_{s's}+\mathcal{V}_q(x_1)\mathcal{\bar V}_q(x_2)f_F(\omega - (m+1/2)\Omega - \mu_a  {\cal R}^a)_{ss'}\right\}, \nonumber\\ 
S_{ss'}^<(x_1,x_2) \equiv -\langle \bar \Psi_{s'}(x_2)\Psi_s(x_1)\rangle  =-\SumInt_q \left\{\mathcal{U}_q(x_1)\mathcal{\bar U}_q(x_2) f_F(\omega - (m+1/2)\Omega - \mu_a \bar {\cal R}^a)_{s's}+\mathcal{V}_q(x_1)\mathcal{\bar V}_q(x_2)(1-f_F(\omega - (m+1/2)\Omega - \mu_a  {\cal R}^a))_{ss'}\right\} \nonumber \eea
that gives us the thermal propagator through Eq.~(\ref{Tpropf}). %However, in computing particle decays or production, it is typically easier to work with the non-time-ordered 2-point functions. 
Also, note that in the case of massless fermions that have just one helicity state, only one term in the sum over helicities in the expressions above should be considered.

Let us now consider the case of Majorana fermions, which are described here by Weyl spinors,  Eq.~(\ref{FreeWexp}). Then we can consider the following types of propagators
 \bea  \langle {\cal T} \Psi_s(x_1)\bar \Psi_{s'}(x_2)\rangle &=& \theta(t_1-t_2)\langle \Psi_s(x_1)\bar \Psi_{s'}(x_2)\rangle  -\theta(t_2-t_1)  \langle \bar \Psi_{s'}(x_2)\Psi_s(x_1)\rangle,  \label{TpropfW1} \\ 
 \langle {\cal T} \Psi_s(x_1)\Psi_{s'}(x_2)\rangle &=& \theta(t_1-t_2)\langle \Psi_s(x_1)  \Psi_{s'}(x_2)\rangle  -\theta(t_2-t_1)  \langle \Psi_{s'}(x_2)\Psi_s(x_1)\rangle, \label{TpropfW2} \eea
where the spinor indices are understood. 
The field expansion in~(\ref{FreeWexp}) then leads to  (recalling $\mathcal{\bar R}^a=\mathcal{R}^a$ for Majorana fermions)
\bea 
&&\hspace{-0.5cm}S_{ss'}^>(x_1,x_2)  \equiv \langle \Psi_s(x_1)\bar \Psi_{s'}(x_2)\rangle= \SumInt_q \left\{X_q(x_1)\bar X_q(x_2)(1-f_F(\omega - (m+1/2)\Omega - \mu_a  {\cal R}^a))_{s's}+Y_q(x_1)\bar Y_q(x_2)f_F(\omega - (m+1/2)\Omega - \mu_a  {\cal R}^a)_{ss'}\right\}, \nonumber\\ 
&&\hspace{-.5cm}S_{ss'}^<(x_1,x_2) \equiv -\langle \bar \Psi_{s'}(x_2)\Psi_s(x_1)\rangle=-\SumInt_q \left\{X_q(x_1)\bar X_q(x_2) f_F(\omega - (m+1/2)\Omega - \mu_a  {\cal R}^a)_{s's}+Y_q(x_1)\bar Y_q(x_2)(1-f_F(\omega - (m+1/2)\Omega - \mu_a  {\cal R}^a))_{ss'}\right\}, \nonumber \\ 
&&\hspace{-.5cm}\tilde S_{ss'}^>(x_1,x_2)  \equiv \langle \Psi_s(x_1)  \Psi_{s'}(x_2)\rangle= \SumInt_q \left\{X_q(x_1)Y_q(x_2)(1-f_F(\omega - (m+1/2)\Omega - \mu_a  {\cal R}^a))_{s's}+Y_q(x_1)X_q(x_2)f_F(\omega - (m+1/2)\Omega - \mu_a  {\cal R}^a)_{ss'}\right\},\nonumber\eea
while $\tilde S_{ss'}^<(x_1,x_2)  \equiv -\langle \Psi_{s'}(x_2)  \Psi_{s}(x_1)\rangle$ 
can easily be obtained from $\tilde S_{ss'}^>(x_1,x_2)$ by exchanging $s\leftrightarrow s'$ and $x_1\leftrightarrow x_2$. 
Analogously, one can also determine $\langle {\cal T} \bar\Psi_s(x_1) \Psi_{s'}(x_2)\rangle$ and $\langle {\cal T} \bar\Psi_s(x_1)\bar\Psi_{s'}(x_2)\rangle$.

One can combine the non-time-ordered 2-point functions to obtain the thermal propagators in~(\ref{Tpropf}),~(\ref{TpropfW1}) and~(\ref{TpropfW2}) expressed in terms of an integral over four momentum variables  $\{k_0, \alpha, p, m\}$ by using the procedure around Eq.~(\ref{4Mom}).

Also for fermion fields \citet{Vilenkin1980} discussed a different 2-point function: the thermal average of the time-ordered product (at different spacetime points) of the Matsubara field operator for vanishing $\mu_a$. This does not coincide with $\Psi_s$ but with the field $\tilde \Psi_s$ (recall the general definition in~(\ref{tildeO})), after the substitution  $it\to \tau$, where $\tau$ is the imaginary time. 

The expressions of the thermal propagators given in this section, together with the information that the vertices do not depend on $\Omega$ and the $\mu_a$, has been applied to neutrino production processes in neutron stars~\citep{Salvio2025b}.

\subsection{Path integral in gauge theories}\label{GaugePI}

Let us now analyze gauge theories to include vector fields. Here we also keep a general value not only of $\Omega$ but also of the chemical potentials $\mu_a$ and some of them can correspond to gauge charges.

We first start from the classical  theory and proceed afterwards to quantization. After all, the most famous gauge theory, electrodynamics, was first known as a classical theory and was quantized only afterwards.
 
  We consider, however,  a general Yang-Mills (not necessarily Abelian) theory, where the gauge group $G$ is a Lie group  with generators $\mathcal{T}^a$, with $a=1, ... , \dim G$. The global subgroup of $G$ can be viewed as a subgroup of the full global group $\mathcal{G}$ and one can again assign chemical potentials to the charges $Q^a$, with $a=1, ... , \dim \mathcal{G}$. We choose the $\mathcal{T}^a$ such that the real constants $f^{abc}$ (the structure constants), defined by
\be [\mathcal{T}^a,\mathcal{T}^b] = i f^{abc} \mathcal{T}^c, \ee
are totally antisymmetric.
  The field strength $F^a_{\mu\nu}$ is given in terms of the gauge fields $A^a_\mu$ by
\be F^a_{\mu\nu} = \partial_\mu A^a_\nu-\partial_\nu A^a_\mu -g f^{bca} A^b_\mu A^c_\nu,\ee
where $g$ represents the gauge couplings. Here a compact notation is used and an index on $g$ is implied, so also the cases with more than one gauge couplings are covered.

We start from the gauge-invariant  classical Lagrangian given by
\be \mathscr{L}= -\frac14 F^a_{\mu\nu}F^{a\mu\nu} + \mathscr{L}_M(\Theta,D\Theta), \ee
where $\mathscr{L}_M$ represents the matter Lagrangian, which we assume here to be an ordinary function of only the matter fields $\Theta$ (which can be scalars and fermions) and their covariant derivatives $D_\mu\Theta$, 
\be D_\mu\Theta=(\partial_\mu+i g\mathcal{T}^a A^a_\mu)\Theta.\ee  

To exploit the cylindrical symmetry induced by $\vec\Omega$ let us choose the axial gauge 
\be A_3^a = 0. \ee
Then the canonical variables of the gauge sector are the $A_i^a$, with $i=1,2$, and the corresponding conjugate momenta read
\be \Pi_a^{i} \equiv \frac{\partial\mathscr{L}}{\partial \dot  A_i^a}  = F^{a}_{0i}, \ee
while $A_0^a$ is not an independent variable, but is a functional of the other variables: such functional is obtained by solving the field equations of $A_0^a$ itself, which represent 
secondary constraints~\citep{Weinberg2013}. 

One can now express a generic thermal Green's function of the form in~(\ref{GreenO}),
   where now the $\mathcal{O}_i$ are generic local    operators involving gauge fields, in terms of a path integral over  $A_i^a$  and $\Pi_a^{i}$. Combining the derivation of such path integral in purely scalar theories, but with a general equilibrium density matrix, of Sec.~\ref{ScalarPI} with that in a non-statistical gauge theory provided by~\citet{Weinberg2013}, we find
\be   \boxed{\langle {\cal T}\mathcal{O}_1(x_1) ...  \mathcal{O}_n(x_n) \rangle =\frac1{Z}\int \delta A_1\delta A_2\, \delta \Pi^1\delta \Pi^2  \, O^\omega_1(x_1^\omega) ... O_n^\omega(x_n^\omega)\exp\left(i \int_C d^4x \left(\dot A_i(x)\Pi^i(x) - \mathcal{H}^\omega_c(A(x). \Pi(x))\right)\right). } \label{PIgauge} \ee
   Some definitions and clarifications are now in order for this result.
   \begin{itemize}
\item The path integration is on all field configurations satisfying  the twisted periodicity condition:
  \be e^{\beta\mu_a f^a}A(t_0,\vec{x}) = A(t_0-i\beta,\vec{x}),\label{PerConA}\ee
  where the $f^a$ are the generators of $\mathcal{G}$ in the  representation of the gauge fields (the adjoint representation), which acts on the Lie-group index.  
  \item The integration measures on the $A_i^a$ and the $\Pi^i_a$ are
  \be \delta A_i = \prod_{x, a} dA^a_i(x), \quad \delta \Pi^i= \prod_{x, a} d(\Pi^i_a(x)/(2\pi)). \label{measureAPI} 
 \ee
\item The function $\mathcal{H}^\omega_c$ is still given by~(\ref{callHc}). In a Yang-Mills theory 
   $H_c$ and $\vec J_c$ read 
\be H_c(A, \Pi)\equiv \frac{\langle A|H|\Pi\rangle}{\langle A|\Pi\rangle}, \qquad \vec J_c(A, \Pi)\equiv \frac{\langle A|\vec J |\Pi\rangle}{\langle A|\Pi\rangle},    \ee 
where $|A\rangle$ and $|\Pi\rangle$ are the generic eigenstates of the canonical  quantum variables. The quantum Hamiltonian $H$ is chosen in a way that   $\mathcal{H}_c$ reproduces  the known expression for the classical Hamiltonian density of the Yang-Mills theory in the absence of rotation: e.g.~in the axial gauge we are using      
\be\mathcal{H}_c= \Pi_a^{i} \left(\partial_i A_0^a+gf^{bca} A^b_0 A^c_i\right)+\frac{1}{2}\Pi_a^{i} \Pi_a^{i}  +\frac14 F^a_{ij}F^{a}_{ij}+\frac12 \partial_3 A^a_i\partial_3 A^a_i-\frac12 \partial_3 A^a_0\partial_3 A^a_0.\label{Hcgauge} \ee
with $A_0^a$ set equal to the solution of its  field equations. Then, $A_0^a$ generically turns out to be a complicated (even non-local) functional of the canonical  variables. This complication will be  avoided below with a further path integration.
\item The quantum angular momentum $\vec J$ is chosen in a way that $\vec J_c$ reproduces  the known expression for the classical angular momentum of the Yang-Mills theory:  the components $J_{ck}$, with $k=1,2,3$, of $\vec J_c$ are given by
\be J_{ck} = \epsilon_{lmk} \int d^3 x\, (\Pi^p_a\, x^m\partial_l A_p^a- A^{a}_m \Pi^l_a ). \label{Jcc}\ee 
This formula holds beyond the axial gauge $A_3^a=0$ (if $A_3^a\neq0$ one defines $\Pi^3_a \equiv \frac{\partial\mathscr{L}}{\partial \dot  A_3^a}  = F^{a}_{03}$).  
Eq.~(\ref{Jcc}) can also be written in a vector notation:
\be \vec J_c = \int d^3x \,(-\Pi^k_a\, \vec x \times \vec\nabla A_k^a + \vec A^a \times \vec \Pi_a), \label{Jcvector} \ee 
where $\vec A^a$ and $\vec \Pi_a$ are the three-vectors with components $A^a_k$ and $\Pi_a^k$, respectively.
The first term in~(\ref{Jcvector}) is the orbital angular momentum contribution, while the second one is the spin contribution.
\end{itemize}

\noindent One can now combine~(\ref{Hcgauge}) and~(\ref{Jcc}) to find in the axial gauge
\be\mathcal{H}_c^\omega= \Pi_a^{i}\left[\partial_i A_0^a+gf^{bca} A^b_0 A^c_i +\Omega (\partial_\phi A_i^a+\epsilon_{ij} A^{a}_j)\right]+\frac{1}{2}\Pi_a^{i} \Pi_a^{i}  +\frac14 F^a_{ij}F^{a}_{ij}+\frac12 \partial_3 A^a_i\partial_3 A^a_i-\frac12 \partial_3 A^a_0\partial_3 A^a_0,
\label{HcOmegagauge}\ee
 where $\epsilon_{ij}\equiv\epsilon_{ij3}$.

 The partition function that appears in the denominator in~(\ref{PIgauge}) can be expressed as
  \be \boxed{Z =\int \delta A_1\delta A_2 \, \delta \Pi^1\delta \Pi^2  \,  \exp\left(i \int_C d^4x \left(\dot A_i(x)\Pi^i(x) - \mathcal{H}^\omega_c(A(x), \Pi(x))\right)\right),} \label{GenPartGauge}  \ee
  which is nothing but the path integral in~(\ref{PIgauge}) without $\mathcal{O}_1(x_1) ...  \mathcal{O}_n(x_n)$. Recall that $Z$ can be used to compute the averages of observables, see e.g.~(\ref{avHg}),~(\ref{avJg}) and~(\ref{avQg}). Eq.~(\ref{measureAPI}) specifies the integration measure to compute $Z$ (see e.g.~\cite{Braguta2025} for applications with $\Omega\neq0$).

   Like in scalar-fermion theories, to account for a non-vanishing average angular momentum one has to substitute  
  $H_c$ with $H_c -\vec \Omega\cdot \vec J_c$ in the path integral, which is nothing but the transformation rule of the classical Hamiltonian from an inertial frame to a frame rotating with the angular-velocity vector $\vec \Omega$.   Again, this tells us that we can identify $\vec \Omega$ with the angular-velocity vector of the rotating plasma.

One can  combine in a straightforward way~(\ref{PIgauge}) and~(\ref{GenPartGauge}) with the results in Secs.~\ref{ScalarPI} and~\ref{FermionPI} to obtain a path integral expression for $Z$ and the Green's functions of  the form in~(\ref{GreenO}), in the presence of both gauge fields and matter fields.

Since $A_0^a$ appears at most quadratically in $\mathcal{H}_c^\omega$  and the quadratic piece $-\frac12 \partial_3 A^a_0\partial_3 A^a_0$ has a field-independent coefficient (see Eq.~(\ref{HcOmegagauge})), one can effectively treat  $A_0^a$ as an independent integration variable:
\be  \langle {\cal T}\mathcal{O}_1(x_1) ...  \mathcal{O}_n(x_n) \rangle = \frac1{``O_i\to 1"} \int \delta A_0 \delta A_1\delta A_2 \, \delta \Pi^1\delta \Pi^2  \, O^\omega_1(x_1^\omega) ... O_n^\omega(x_n^\omega)  \exp\left(i \int_C d^4x \left(\dot A_i(x)\Pi^i(x) - \mathcal{H}^\omega_c(A(x), \Pi(x))\right)\right),\label{PIgauge2} \ee
where 
 \be \delta A_0 = \prod_{x, a} dA^a_0(x).\ee
Indeed, path integrating over $A_0^a$ such Gaussian function of  $A_0^a$  sets $A_0^a$ equal to the solution of its  field equations and produces a field-independent coefficient that cancels with the denominator $``O_i\to 1"$ (which coincides with the numerator for $O^\omega_1(x_1^\omega) ... O_n^\omega(x_n^\omega)\to 1$). To see that this is true even if the $O_i$ depend on $A_0^a$, one can first introduce a generating functional  coupling all fields to external currents, including $A_0^a$. The result of the path integration over $A_0^a$ would then sets $A_0^a$ equal to the solution of its  field equations including the external-current contribution. The Green's function are then generated by taking functional derivatives with respect to the external currents and then setting those currents to zero, which sets $A_0^a$ equal to the solution of its  field equations without the external-current.

With $A_0^a$ regarded as an independent integration variable, $\mathcal{H}_c^\omega$ is clearly at most quadratic in the $\Pi^i_a$ with the quadratic piece $\frac{1}{2}\Pi_a^{i} \Pi_a^{i}$ having a field-independent coefficient (see again Eq.~(\ref{HcOmegagauge})). The corresponding Gaussian path integral over $\Pi^i_a$ sets 
\be \Pi^i_a =  F^{a}_{0i} - \Omega (\partial_\phi A_i^a+\epsilon_{ij} A^{a}_j)\ee 
and produces a field-independent coefficient that cancels again with the denominator $``O_i\to 1"$.
This sets the integrand in the exponential in~(\ref{PIgauge2}) equal to the Lagrangian density 
$\mathscr{L}_\omega$, which is obtained by substituting  in  $-\frac14 F^a_{\mu\nu}F^{a\mu\nu}$ \be \partial_0 A_k^a\to\partial_0 A_k^a - \Omega (\partial_\phi A_k^a+\epsilon_{kl3} A^{a}_l) \label{subL}\ee
(for all $k,l=1,2,3$). 
The path integral in~(\ref{PIgauge2}) can then be expressed in the Lagrangian form:
\bea \langle {\cal T}\mathcal{O}_1(x_1) ...  \mathcal{O}_n(x_n) \rangle = \frac1{``O_i\to 1"}&&\hspace{-0.6cm}\int \delta A \, \delta(A_3)  \, O^\omega_1(x_1^\omega) ... O_n^\omega(x_n^\omega) \exp\left(i \int_C d^4x \mathscr{L}_\omega\right),\label{PIgauge3}\hspace{-0.3cm}\eea 
where the functional delta function
\be \delta(A_3) \equiv \prod_{x, a}\delta(A^a_3(x)) \ee
enforces the axial gauge and
 \be \delta A \equiv \delta A_0\delta A_1 \delta A_2 \delta A_3, \qquad \delta A_3 \equiv \prod_{x, a} dA^a_3(x).\ee

Note that, like for the scalars, since going from a non-rotating to a rotating frame corresponds to substituting $H\to H-\vec\Omega \cdot \vec J$, the convergence of the partition function  requires $H-\vec\Omega \cdot \vec J$ (at least for $\mu_a=0$) and the real part of the Euclidean action to be bounded from below. 
Moreover, at least on a finite lattice, the convergence of the ensemble averages of all operators imply the convergence of all Euclidean thermal Green's functions.
 
The same action 
$\int_C d^4x\mathscr{L}_\omega$ can be equivalently obtained by starting from $-\frac14 \int_C d^4x F^a_{\mu\nu}F^{a\mu\nu}$ and changing the gauge four-vector as (recall that the transformation $\vec x\to R(t\vec\Omega)\vec x$ corresponds to $\phi\to\phi-\Omega t$)
\be A^a (t,\vec x) \to \mathscr{R}^{-1}(t\vec\Omega)A^a (t,R(t\vec\Omega)\vec x)\equiv \mathscr{R}^{-1}(t\vec\Omega)A^a (x^\omega), \ee 
where $\mathscr{R}(t\vec \Omega)$ is the $4\times 4$ matrix, which represents the rotation of an angle $t\Omega$ around $\vec\Omega$ on four-vectors.
 To show the equivalence note that the  change of the spatial integration variable in the action from $\vec x$ to $R^{-1}(t\vec \Omega)\vec x$ has unit Jacobian for any (even complex) $t$.
So  one sees again that having a plasma with a non-vanishing thermal vorticity is equivalent to going from an inertial frame to a frame that is rotating with angular-velocity vector $\vec \Omega$. 
This also implies the gauge invariance of $\mathscr{L}_\omega$.  Therefore, if the operators $\mathcal{O}_i$ are also gauge invariant, we can use the Faddeev-Popov argument to rewrite the thermal Green's function in a generic gauge:
 \bea \langle {\cal T}\mathcal{O}_1(x_1) ...  \mathcal{O}_n(x_n) \rangle = \frac1{``O_i\to 1"}&&\hspace{-0.6cm}\int \delta A \, D_f(A) \delta(f(A)) \, O^\omega_1(x_1^\omega) ... O_n^\omega(x_n^\omega) \exp\left(i \int_C d^4x \mathscr{L}_\omega\right),\label{PIgauge4}\hspace{0.9cm}\eea 
where here the function $f$ identifies the generic gauge condition through $f_a(A,x)=0$ (for example, in the axial gauge $f_a(A,x) = A^a_3(x)$), 
\be  \delta(f(A)) \equiv \prod_{x, a}\delta(f_a(A,x)) \ee
and $D_f(A)$ is the determinant of the ``matrix"
\be M_{ax,by}(A) \equiv \left.\frac{\delta f_a(A^\alpha,x)}{\delta \alpha^b(y)}\right|_{\alpha =0}, \ee 
with $A^\alpha$ being the gauge-field configuration  that is obtained from the gauge-field configuration  $A$ through the element of the gauge group with parameters $\alpha^a$ . As usual,  we can generically substitute $D_f(A)$ with an integral over Grassmann field variables, $c^a$ and $\bar c^a$ (Faddeev-Popov ghosts):
\be D_f(A) \to \int \delta\bar c\,  \delta c \, \exp\left(i\int_C d^4x d^4y \, \bar c^a(x) M_{ax,by}(A) c^b(y)\right),  \label{GhostInt}\ee
where 
\be \delta c = \prod_{x, a} dc^a(x), \qquad \delta \bar c = \prod_{x, a} d\bar c^a(x). \ee 
Since the operator associated with $M_{ax,by}(A)$ acts on a space of functions satisfying the twisted periodicity boundary conditions~(\ref{PerConA}), the Faddeev-Popov ghosts obey the same conditions although they are Grassmann variables. Faddeev-Popov ghosts will be interacting with some other fields any time $D_f(A)$ depends on those fields. This is not the case, e.g., in the axial gauge, so  $D_f(A)$ cancels with the denominator $``O_i\to 1"$ in that gauge and~(\ref{PIgauge3}) is recovered. 

We can generate the thermal Green's function by performing functional derivatives of the generating functional 
 \be \boxed{ \mathcal{Z} (J) =  \frac1{``O_i\to 1"} \int \delta A \, D_f(A) \delta(f(A)) \, \exp\left(i \int_C d^4x \left( \mathscr{L}_\omega + J(x)\mathscr{R}^{-1}(t\vec\Omega) A(x^\omega)\right)\right).\label{PIgenGauge}} \ee
 
Now, let us make a remark on the extension of~(\ref{PIgauge4}) to include matter fields. If matter fields are also present  one should also integrate over them with measures $\delta\varphi$ and $\delta\bar\psi\delta\psi$ for scalars and fermions, respectively,  and add to  $\mathscr{L}_\omega$ the matter $\Omega$-dependent Lagrangian density, which was studied in Secs.~\ref{ScalarPI} and~\ref{FermionPI}.

 The axial gauge is very useful to derive the path-integral representation of the thermal Green's functions from first principles. However, now one can choose an arbitrary gauge to perform calculations. 
 Another possible convenient choice  is \be f_a(A,x) = (\partial_0 - \Omega \partial_\phi )A_0^a(x)-\partial_kA_k^a(x) +\tau_g^a(x),\label{tauGauge}\ee
 where $\tau_g^a$ are fixed fields and different choices of them are thus different gauges.  However, $M_{ax,by}(A)$ and $D_f(A)$ are $\tau_g$-independent. Explicitly, with the gauge choice in~(\ref{tauGauge}), the ghost insertion in the path integral reads
\be D_f(A) \to \int \delta\bar c\,  \delta c \, \exp\left(i\int_C d^4x \, \bar c^a(x) \left\{\delta_{ab}\Box_\omega-gf^{abc}\left[(\partial_0 - \Omega \partial_\phi )A_0^c(x)-\partial_kA_k^c(x)\right]\right\} c^b(x)\right),  \label{GhostInttau}\ee
with
\be \Box_\omega  = \left(\partial_0 - \Omega \partial_\phi\right)^2-\partial_k\partial_k,  \ee
the same differential operator appearing in the quadratic scalar-field action in~(\ref{SC}). Both the quadratic part and the interaction terms in the ghost Lagrangian depend on $\Omega$ with the gauge choice in~(\ref{tauGauge}). In Sec.~\ref{Interactions}, however, we will see that in perturbation theory only the propagators are $\Omega$-dependent, the vertices are not and equal those with $\Omega=0$ and $\mu_a=0$.

One can exploit the fact that both the numerator and the denominator in~(\ref{PIgauge4}) are gauge invariant and thus independent of $\tau_g$ and multiply both of them by the path integral over $\tau_g^a$ of $ \exp(-i\int_C d^4x \, \tau_g^a(x)\tau_g^a(x)/(2\xi_g))$. Using then $\delta(f(A))$ to perform the path integral over $\tau_g$ the Lagrangian density of the gauge fields is changed as follows
\be\mathscr{L}_\omega\to \mathscr{L}_\omega -\frac{1}{2\xi_g} \sum_a\left[(\partial_0 - \Omega \partial_\phi )A_0^a(x)-\partial_kA_k^a(x)\right]^2. \label{xiLag} \ee 
For $\xi_g\to0$ one obtains a ``rotating Landau gauge", $(\partial_0 - \Omega \partial_\phi )A_0^a(x)-\partial_kA_k^a(x)=0$,
while  the $\xi_g=1$ choice is a ``rotating Feynman gauge", which leads to some simplification in perturbation theory, as shown in Sec.~\ref{propGauge}.

\subsection{Gauge thermal propagators} \label{propGauge}

As explained at the end of Sec.~\ref{GaugePI}, the Lagrangian density changes by changing the gauge in a generically $\Omega$-dependent way (see, for example,~Eq.~(\ref{xiLag})). This implies not only that the thermal propagators of the gauge fields can depend on the gauge, but also that some extra gauge-dependent $\Omega$ dependence can appear. The possible ways of fixing the gauge are infinite, so here we focus on a particularly simple choice, the rotating Feynman gauge ($\xi_g=1$ in Eq.~(\ref{xiLag})): as we shall see here, this setting indeed leads to a simple form of the gauge-field thermal propagators that can in turn simplify the perturbative expansion.

In the rotating Feynman gauge the quadratic Lagrangian density for gauge fields turns out to be
\be \frac12 A_0^a\Box_\omega A_0^a-\frac12 A_3^a\Box_\omega A_3^a-\frac12 A_i^a\Box_\omega A_i^a, 
 \ee
which is obtained by considering only the terms quadratic in the gauge fields in the right-hand side of~(\ref{xiLag}) for $\xi_g=1$. The differential operator $\Box_\omega$ acts on $A_0^a$ and $A_3^a$ like on scalar fields: it can be obtained from the d'Alembertian operator $\Box$ through the substitution $\partial_t\to \partial_t-\Omega\partial_\phi$. However,  $\Box_\omega$ acts on $A_i^a$ differently: its action on $A_i^a$ can be obtained from  $\Box A_i^a$ through the more involved substitution $\partial_tA_i^a\to \partial_tA_i^a-\Omega\partial_\phi A_i^a -\Omega \epsilon_{ij}A_j^a$. So the quadratic Lagrangian of $A_3^a$ (and $A_0^a$) is equal (respectively opposite) to that of massless scalars, while that of $A_i^a$ is more complicated due to its non-invariance under rotations around $\vec\Omega$.

Looking at~(\ref{PIgenGauge}), the coupling of $A_0$ and $A_3$ to the external current is the same as that of scalars. As a result, the thermal propagator of $A_3$ (and $A_0$) in this gauge coincides with (is respectively opposite to) that of a massless scalar in the adjoint representation of the gauge group, which is a particular case of the setup already analyzed in Sec.~\ref{ScalarProp}. 

 On the other hand,  the coupling of $A_i$ to the external current in the action is
\be \int_Cd^4xJ^i(x)R_{ij}(-t\vec\Omega) A_j(x^\omega)=\int_Cd^4xR_{ji}(t\vec\Omega)J^i(x^{-\omega}) A_j(x), \ee
where a rotation of angle $t\Omega$ around $\vec\Omega$ of the spatial integration variables was performed. Note that this coupling, unlike the analogous coupling for scalars in~(\ref{ChangSpI}), includes the rotation matrix $R(t\vec\Omega)$. Adapting the procedure presented for scalar fields in Sec.~\ref{ScalarProp}, one finds that the presence of the rotation matrix in the coupling to the external current precisely cancels the presence of $\Omega$ in the equation that the thermal propagator satisfies, which then reads 
\be i\Box^x \langle {\cal T}A_i^a(x)A_j^b(y)\rangle  = \delta_{ij}\delta_{ab}\delta(x-y).\label{EqGFDo2} \ee 
Just like for scalars, that equation does not depend on the $\mu_a$ and $\Omega$. This dependence  will be introduced again by the generalized KMS conditions in~(\ref{KMSc}), which are, however, different for vector operators because of the presence of the matrix $\mathcal{D}_1(-i\beta\vec\Omega)$. For the $A_i$ this matrix is
\be \mathcal{D}_1(-i\beta\vec\Omega) = e^{\beta \Omega\sigma_2 }, \ee
where $\sigma_2$ is the second Pauli matrix. The KMS conditions can then be imposed in a way similar to that presented in Sec.~\ref{ScalarProp} for scalars, except that one should take into account this non-trivial matrix. The result for $G_{ij}^{>ab}(x_1,x_2)\equiv  \langle A_i^a(x_1)A_j^b(x_2)\rangle$ is
\bea G^>(x_1, x_2) &=& \sum_{m=-\infty}^{+\infty}\int_0^\infty d\omega\int_{-\omega}^{\omega}dp \left[ (1+f_B (\omega-(m+\sigma_2)\Omega+\mu_af^a))e^{-i\omega t+ipz+im\phi} \frac{J_m(\alpha r_1)J_m(\alpha r_2)}{2^3\pi^2}\right. \nonumber \\
&&\left.+ f_B (\omega-(m-\sigma_2)\Omega-\mu_af^a)e^{i\omega t-ipz-im\phi} \frac{J_m(\alpha r_1)J_m(\alpha r_2)}{2^3\pi^2}\right].\label{GFprop}\eea
 The other non-time-ordered 2-point function  can be easily obtained by using the property $G_{ij}^{<ab}(x_1,x_2) = G_{ji}^{>ba}(x_2,x_1)$ and one can then easily combine $G^>$ and $G^<$ to obtain the thermal propagator: $$\langle {\cal T}A_i^a(x_1)A_j^b(x_2)\rangle=\theta(t_1-t_2)G_{ij}^{>ab}(x_1,x_2)+\theta(t_2-t_1)G_{ij}^{<ab}(x_1,x_2).$$ 
The thermal propagator above can be expressed in terms of an integral over four momentum variables  $\{k_0, \alpha, p, m\}$ by using the procedure around Eq.~(\ref{4Mom})  %XXXXXX{\color{red}
for $t_1$ and $t_2$ on the real axis.
%}.
%.

One can diagonalize $\sigma_2$ in~(\ref{GFprop}) by going  to the spin basis $A_\pm^a\equiv (A_1^a\pm iA_2^a)/\sqrt{2}$, where $ \langle A_\pm^a(x_1)A_\pm^b(x_2)\rangle = 0$ and $ \langle A_\pm^a(x_1)A_\mp^b(x_2)\rangle$ reads as~(\ref{GFprop}), except that $\sigma_2$ is substituted by $\mp 1$, respectively. Once the gauge group is specified, one can then compute $f_B$ at the matrices $\omega-(m\mp 1)\Omega+\mu_af^a$ and $\omega-(m\pm 1)\Omega-\mu_af^a$ by determining  the eigenvalues and the projectors of $\mu_af^a$   as explained around Eq.~(\ref{SpDec}).  

The thermal propagator of the Faddeev-Popov ghosts is also gauge dependent. Let us consider now the gauges~(\ref{tauGauge}), which include the $\xi_g$ gauges (and thus the Feynman gauge) as particular cases.
In those gauges, the differential operator appearing in the piece of the Lagrangian that is quadratic in the Faddeev-Popov Ghosts is the opposite of $\Box_\omega$ for scalars (see~(\ref{GhostInttau})). Furthermore,  the Faddeev-Popov ghosts  obey  the same twisted periodicity boundary conditions of gauge fields. As a result, the ghost thermal propagator in the gauges~(\ref{tauGauge}) is minus  the thermal propagator  of a massless scalar  in the adjoint representation of the gauge group. The same is true for the two non-time-ordered 2-point functions.

\subsection{Interactions}\label{Interactions}

The substitution in~(\ref{subL}) shows that  in a non-Abelian Yang-Mills theory $\Omega$
  affects not only the part of the action quadratic in the field variables, but also introduces the following additional cubic interaction in the Lagrangian density:
 \be \Omega gf^{bca}A^b_0 A^c_k(\partial_\phi A_k^a +\epsilon_{kl 3} A^a_l). \label{extraIntNA}\ee 

Let us see now if~(\ref{extraIntNA}) affects perturbation theory. To this purpose note that the substitution in~(\ref{subL}) also shows that the term~(\ref{extraIntNA}) is  accompanied by another term with $\partial_\phi A_k^a +\epsilon_{kl 3} A^a_l$ substituted by $-\partial_0 A_k^a$. This is the only interaction term involving the time-derivative of the gauge fields. One  can express both of them in terms of  $\mathscr{R}^{-1}(t\vec\Omega) A^a(x^\omega)$ and perform a change of variables (a rotation of angle $t\Omega$ around $\vec\Omega$, which has unit Jacobian) 
  in the spatial integration,  in order to calculate in the perturbative expansion   
  the
 correlators with field insertions of the form $O^\omega(x_1^\omega)...O^\omega(x_n^\omega)$. But doing so one finds that the $\Omega$ dependence in the interactions disappears there.  
So   although $\Omega$ affects the propagators no $\Omega$-dependent vertices are introduced in perturbation theory, like in the general fermion-scalar theories of Secs.~\ref{ScalarPI} and~\ref{FermionPI}.
 
 The (gauge-dependent) interactions of the Faddeev-Popov ghosts with the other fields may also depend on $\Omega$. To illustrate this point let us consider the gauges~(\ref{tauGauge}) (including the $\xi_g$  gauges and thus the Feynman gauge). From~(\ref{GhostInttau}) one  sees that, expressing the interactions of the Faddeev-Popov ghosts in terms of \{$\mathscr{R}^{-1}(t\vec\Omega) A^a(x^\omega)$, $c^b(x^\omega)$, $\bar c^a(x^\omega)$\} 
and performing the above-mentioned change of variable in the spatial integration, again the $\Omega$ dependence  disappears in the interactions. So in perturbation theory there are no $\Omega$-dependent vertices involving Faddeev-Popov ghosts either. 

Finally, let us see  if/how $\Omega$ affects interactions in the matter sector of a gauge theory. As shown in Secs.~\ref{ScalarPI} and~\ref{FermionPI}, $\Omega$ appears only in the quadratic part of the matter Lagrangian of a pure scalar-fermion theory. However, introducing gauge fields, there is one (and only one) $\Omega$-dependent interaction between gauge fields and matter, which appears in the scalar sector: this comes from the gauge kinetic terms of the scalars $\varphi$  in the Lagrangian density
\be \frac12 [(\partial_0-\Omega\partial_\phi +ig\theta^a A^a_0)\varphi](\partial_0-\Omega\partial_\phi +ig\theta^a A^a_0) \varphi.\label{gaugeKinS}\ee
Again one can see that this $\Omega$-dependent interaction produces no $\Omega$-dependent vertices in perturbation theory. To  see this one can express~(\ref{gaugeKinS})   in terms of $\mathscr{R}^{-1}(t\vec\Omega) A^a(x^\omega)$ and $\varphi(x^\omega)$ and perform again the above-mentioned change of variable in the spatial integration.

So,  in perturbation theory, only the propagators are modified by $\Omega$, the vertices are unmodified in a general gauge theory with an arbitrary matter sector.  For the computation of the vertices one can use well-known results from the literature (see again e.g.~\citet{Bellac2011} and~\cite{Landsman1987}); on the other hand, the propagators have been computed, including the effect of $\Omega$ and generic $\mu_a$,  in Secs.~\ref{ScalarProp}  (for scalars),~\ref {Fermion thermal propagator} (for fermions) and~\ref{propGauge} (for gauge fields and Faddeev-Popov ghosts). This provides a practical and model-independent recipe to perform perturbation theory in the presence of both $\Omega$ and  $\mu_a$. 

For example, one can compute interaction rates (including decay and production rates) of a particle, which is weakly-coupled to the thermal plasma, with the Kobes-Semenoff rules~\citep{Kobes1985}-\citep{Kobes1986}, an extension to TFT of the well-known cutting rules in non-statistical field theory.  Kobes and Semenoff assumed $\Omega = 0$ and $\mu_a = 0$, but, as shown above, only the propagators are modified by $\Omega$ and $\mu_a$, the vertices are unmodified. For example, to  compute the production rate of a spin-1 particle, which is weakly-coupled to the thermal plasma, one can use the formul\ae~in Sec.~5.2 of I1, but without relying on the invariance under space translations perpendicular to $\vec\Omega$ and using the decomposition of plane waves in cylindrical Bessel functions. On the other hand, some applications to fermion production for $\Omega\neq0$ were provided by~\cite{Castano2025} and~\cite{Salvio2025c}.

\section{Summary and conclusions}\label{Summary and conclusions}
This chapter provided the general tools to perform computations in a  TFT with arbitrary equilibrium density matrices, which feature arbitrary values of the temperature, the chemical potentials and the average angular momentum. The approach we followed was fully model-independent, so one can apply the results presented here to any specific field-theory model of physical relevance. Indeed, scalars, fermions (both of Dirac and Majorana type) and vectors were all covered. 

We first described free particles, explicitly showing in this simple case how to compute the partition function and the ensemble averages of the relevant quantities, including the Hamiltonian, the angular momentum and the conserved charges generating the internal-symmetry group. 

Next we switched on general interactions and (1) showed the form of the KMS conditions in this general setup and (2) derived path-integral representations for the partition function and the Green's functions of an arbitrary number of general local operators.
In doing so we considered a generic gauge theory, which can be coupled to an arbitrary matter sector made of scalars and fermions. A formalism that includes the imaginary-time and real-time formalism as specific choices was used. So one can apply the results presented here both in perturbative and non-perturbative calculations. 

Regarding perturbation theory, the thermal propagators were derived in all sectors of the theory (for scalars, fermions and in the gauge sector). Furthermore, we have learnt that the vertices are not affected by the rotation of the plasma and the chemical potentials. This provides us with a practical recipe to perform perturbation theory whenever the perturbative expansion gives good approximations.

\begin{ack}[Acknowledgments]
~ I thank Francesco Tombesi for valuable discussions on rotating plasmas around black holes.
\end{ack}

%\seealso{article title article title}

%\bibliographystyle{Numbered-Style}
%\bibliography{MRW-CQUA-reference}

%\bibliography{references}

\end{document}